%% file: optobs.tex
\documentclass{article}

\usepackage{graphicx,amsmath,amssymb,subfigure,bbm,epstopdf}
\usepackage{color}
\usepackage[sort&compress,numbers]{natbib}
\usepackage{soul}
\usepackage{authblk}
\usepackage{caption,url,soul}
\usepackage[makeroom]{cancel}

\input{sandu_template_symbols}

\graphicspath{ {./Figures/} }

\begin{document}

\include{logo}

\title{An Optimization Framework to Improve 4D-Var Data Assimilation System Performance}

\date{\today}

\author{Alexandru Cioaca}
\author{Adrian Sandu}

\affil{Computational Science Laboratory \\ Department of Computer Science \\
Virginia Tech \\
2202 Kraft Drive, Blacksburg,  Virginia 24060 \\
\tt{alexgc@vt.edu}, 
\tt{asandu7@vt.edu}
}

\maketitle

\begin{abstract}

This paper develops  a computational framework for optimizing the parameters
of data assimilation systems in order to improve their performance.
The approach formulates a continuous meta-optimization 
problem for parameters; the meta-optimization is constrained by the original data assimilation problem.
The numerical solution process
employs adjoint models and iterative solvers. The proposed framework is
applied to optimize observation values, 
data weighting coefficients, and the location of sensors for a test problem.
The ability to optimize a distributed measurement network is crucial
for cutting down operating costs and detecting malfunctions.

\end{abstract}

\tableofcontents
\newpage


\section{Introduction}

Data assimilation improves our understanding of natural phenomena by
integrating measurements of the physical system with computer-generated predictions.
Important theoretical and practical questions
regarding the optimal strategies for collecting the data, and for incorporating it
 such as to obtain the maximum informational benefit, are yet to be answered. 
This paper develops a computational framework for improving the performance of
data assimilation systems by optimally configuring  various parameters of the measuring network.
The methodology developed herein has applicability to many other inverse problems and dynamic data driven applications.

The research is carried out in the context of 
``four-dimensional variational'' (4D-Var) data assimilation, arguably the most advanced framework for assimilating observations 
distributed in time and space with predictions performed with nonlinear numerical models.
4D-Var is extensively used in meteorology, climatology, hydrology and other environmental studies \cite{Kalnay,Daley,lewisds,navon2009data,chai2007four}.
In the 4D-Var approach the data assimilation problem is posed as a PDE-constrained nonlinear \textit{optimization} problem,
where model states and parameters are tuned in order to obtain predictions that fit best the measurements.
{\em In this work the strategies used to collect and process the data are considered to be parameters of the 4D-Var data assimilation system,
and are themselves improved via an additional optimization process.} 

Previous related research has been motivated by the need to quantify the contribution of various 4D-Var parameters on the performance of the system.
The fastest directions of state error growth have been identified with the help of singular vectors \cite{Palmer_1999, Sandu_HSV}.
In order to quantify the amount of information carried by each individual data point, 
information theory \cite{ZupInfoTheory,singh2013practical} or statistical design \cite{Berliner_1999} methodologies have been used.
An adjoint-sensitivity analysis approach has been employed to assess the contributions of individual observations to reducing forecast errors (observation impact)  \cite{gelaro2007examination,gelaro2009examination,TELA:TELA349}.
These methods have been used to qualitatively guide targeting strategies for adaptive observations \cite{baker2000observation,daescu2004adaptive}.
The optimal configuration of observing networks has proved to be very challenging challenging even for the simpler problem of state-estimation \cite{ucinski2000optimal}.
For data assimilation, optimal sensor configuration problem was only partially solved using concepts from control theory such as observability \cite{TELLUSA17133}.
Observing system simulation experiments \cite{wang2008hybrid,atlas1997atmospheric} offer an ad-hoc solution by searching for good network configurations though the entire set of all possible ones.
 
This research advances the current state of science by developing  a systematic approach to tune various 4D-Var parameters
in order to improve the performance of the data assimilation system.
The proposed framework defines a ``meta''-\textit{optimization} problem on top of the 4D-Var solution.
The meta-problem seeks finds the optimal parameter values with respect to a performance functional,
and is constrained by another optimization problem -- the 4D-Var.  The numerical solution process utilizes the
4D-Var sensitivity equations \cite{LeDimet_1997,Daescu_2008,Daescu_2010}; to the best of our knowledge the present work is
the first study to employ this approach. We illustrate the application of the proposed framework
to find optimal observation \textit{values}, observation \textit{weights}, and sensor \textit{locations}.
The proposed framework is general and can be employed to meta-optimize any other parameters of a 4D-Var
data assimilation system. 
  
The paper is structured as follows. Section \ref{sec:optobs_da} reviews the formulation 
of the 4D-Var data assimilation problem, and the associated sensitivity equations.
Section \ref{sec:optobs_contopt} constructs the new optimization based framework for improving the data assimilation system.
Numerical results obtained with the two-dimensional shallow water equations illustrate the applicability of the framework in Section \ref{sec:optobs_appl}. 
Conclusions and directions for future research are provided in Section \ref{sec:optobs_concl}.

\section{4D-Var Data Assimilation}\label{sec:optobs_da}

In this section we review the 4D-Var data assimilation problem formulation 
\cite{Daley,Kalnay_2002,cacuci1981sensitivity,Wang_1992,sandu2008discrete}, 
and the sensitivity equations of the 4D-Var solution
with respect to various system parameters \cite{Daescu_2008}.
These elements provide the building blocks for optimizing the data assimilation system,
including the sensor network configuration.

\subsection{Formulation}

Data assimilation (DA) combines three sources of information: a priori estimates on the system states (``background''),
knowledge of the governing physical laws (``model''), and measurements of the real system (``observations'').
The three sources of information are reconciled through a series of computational steps
in order to generate improved (``analyzed'') estimates of the model states or model parameters.

4D-Var data assimilation takes a Bayesian approach and seeks maximum likelihood estimates of these values with respect to the 
posterior probability density (conditioned by observations).
 4D-Var assimilation is formulated as a PDE-constrained nonlinear optimization problem where
the analysis (the most likely initial state) $\xa_0$ is obtained by minimizing the following (negative log-likelihood) cost function:
 \begin{subequations}
 \begin{eqnarray}
  \label{eqn:optobs_fdvar-costfun}
  \Jfunc(\x_0) &=& \frac{1}{2} \left( \x_0 - \xb_0 \right)^T \cdot \B_0^{-1} \cdot ( \x_0 - \xb_0 )   \\
             & & \, + \frac{1}{2} \sum_{k=0}^{N} \left( \Hobs_k (\x_k) - \y_k \right)^T \cdot \R_k^{-1} \cdot \left( \Hobs_k (\x_k) - \y_k \right)\,, 
  \nonumber \\
 \label{eqn:optobs_fdvar-optimization}  
 \xa_0 &=& \arg\min_{\x_0} \Jfunc(\x_0) \quad \textnormal{subject to }\x_k = \Model_{t_0 \rightarrow t_k} (\x_0)\,.
 \end{eqnarray}
 \end{subequations}
Here $\Model_{t_0 \rightarrow t_k}$ represents the numerical model used to evolve the initial state vector $\x_0$ to future times $t_k$, $k=0,\dots,N$. 
This is called the {\em forward model} ({\sc fwd}) or forecast model. 
$\Hobs_k$ is the observation operator 
which maps the model state $\x_k \approx \x(t_k)$ onto the observation space. 
The error covariance matrices $\B_0$ and $\R_k$ quantify the uncertainty in the background state ($\xb_0$) and in the observations
($\y_k$ for each $t_k$), respectively; they are prescribed by the user and their choice influences the quality of the resulting analysis.

The name ``four-dimensional'' indicates that the method operates with time-distributed observations.
When assimilating observations only at the initial time $t_0$, the method is known as ``three-dimensional variational''
(3D-Var), as the additional time dimension is not present.

\subsection{Numerical solution}

The analysis (initial) state $\xa_0$ is the unconstrained minimizer of \eqref{eqn:optobs_fdvar-optimization} and satisfies the first-order optimality condition:
\begin{equation}
\nabla_{\x_0}\, \Jfunc(\xa_0) = \B_0^{-1} \left(\xa_0 - \xb_0\right) + \sum_{k=0}^{N}  \M_{0,k}^T \HH_k^T \R_k^{-1} \left( \Hobs_k(\xa_k) - \y_k \right) = 0 \,,
\label{eqn:optops_fdvar-fooc}
\end{equation}
where $\M_{0,k} = \Model_{t_0 \rightarrow t_k}'(\x_0)$ is the tangent linear propagator associated with the 
numerical model $\Model$, while $\M^T_{0,k}$ represets its adjoint counterpart.
$\HH_k=\Hobs_k'(\x_k)$ is the linearized observation operator at time $t_k$. 

The minimizer $\xa_0$ is computed numerically using gradient-based methods such as
quasi-Newton \cite{dennis1977quasi}, nonlinear conjugate gradients \cite{dai1999nonlinear}, or truncated Newton \cite{schlick1992tnpack}.
These methods require the derivatives of $\Jfunc$ to the initial model states $\x_0$,
which can be computed using the methodology of adjoint models \cite{cacuci1981sensitivity,Wang_1992}.
Adjoint models have been successfully implemented in optimization, sensitivity analysis and uncertainty quantification
\cite{sandu2008discrete,SanduADJ_2005,Cioaca_2011}.
In previous research we have constructed adjoint models of order up to two and have
compare the performance of different gradient-based optimization algorithm for solving \eqref{eqn:optobs_fdvar-optimization} \cite{Cioaca_2011}.

\subsection{Sensitivity equations}

Using the framework of sensitivity analysis and adjoint models, one can compute 
how small changes in the data assimilation parameters translate into changes in the resulting analysis.
The derivation below follows the 4D-Var sensitivity approach of Daescu \cite{Daescu_2008}.

Consider the 4D-Var problem \eqref{eqn:optobs_fdvar-optimization} where cost function $\Jfunc$
depends on a vector of parameters $\u \in \mathbb{R}^m$:
\[
 \xa_0(\u) = \arg\min_{\x_0} \Jfunc(\x_0,\u) \quad \textnormal{subject to }\x_k = \Model_{t_0 \rightarrow t_k} (\x_0,\u)\,.
 \]
The optimal solution $\xa_0$ satisfies the first order optimality condition for any $\bar{\u}$:
\begin{align}
 \nabla_{\x_0} \, \Jfunc\left(\xa_0(\bar{\u}),\bar{\u}\right) = 0 \,.
\label{eqn:optobs_fooc}
\end{align}
The 4D-Var Hessian is positive definite at the optimum, $\nabla_{\x_0,\x_0}^2 \Jfunc(\xa_0(\bar{\u}),\bar{\u})>0$.
The \textit{implicit function theorem} \cite{krantz2002implicit} applied to \eqref{eqn:optobs_fooc} 
guarantees that there exists a neighborhood of $\bar{\u}$ where the optimal solution $\xa_0(\u)$ is 
a smooth function of the parameters, and 
\begin{align}
\label{eqn:dxa-du}
 \nabla_\u\, \xa_0(\u) = -\nabla_{\u,\x_0}^2 \Jfunc (\xa_0(\u), \u) \cdot \left(\nabla_{\x_0,\x_0}^2 \Jfunc(\xa_0(\u), \u)\right) ^{-1}\,. 
\end{align}
The gradient of the analysis to the set of parameters is the negative product
between the gradient of the first-order optimality condition with respect to the set of parameters,
and the inverse of the 4D-Var Hessian with respect to model states. 


In our previous research we have developed methods to compute efficiently the sensitivity to observations
and the ``observation impact'' using second-order adjoint models, preconditioners \cite{Cioaca_2012}, and
low-rank approximations based on the singular value decomposition \cite{Cioaca_2013}. 

A complete set of sensitivity equations for observations, observation covariances, background and
background covariances can be found in the original derivation of Daescu \cite{Daescu_2008}.
Each of the three applications presented later in this paper requires a different sensitivity equation,
which is provided along with the description of the problem.

\section{Improving the 4D-Var Data Assimilation System via Continuous Optimization}\label{sec:optobs_contopt}

This section develops a methodology to improve the performance of the 4D-Var data assimilation system  
by optimizing the system parameters with respect to a certain objective.
Our specific goal in this paper is to enhance the observing network, but other aspects of the system can be 
treated in a similar manner.
The proposed \textit{continuous optimization} approach is general, elegant, and easy to implement,
since it reuses the same adjoint models and numerical libraries employed for solving the 4D-Var problem.

\subsection{General formulation of the parameter optimization problem}

The first step in defining the optimization problem is to choose the aspects of the data assimilation system that one seeks to improve.
The objective function is a metric defined on the analysis (the output of the assimilation system).
The optimization process finds the values of a set of system parameters 
which minimize (or maximize) the chosen metric.
In a continuous optimization setting the objective function must be  defined 
over a continuous space, where additional constraints can also be imposed.

Many data assimilation studies \cite{daescu2004adaptive,gelaro2007examination,Daescu_2008}
quantify the performance of the 4D-Var system  by the discrepancy between the
model forecast (initialized from the analysis $\xa_0$) at verification time $t_v$ 
and a verification forecast $\x^{\rm verif}_v$, defined at the same time.
The verification $\x^{\rm verif}_v$ represents a control (reference) estimate of the true state.
The magnitude of this discrepancy can be measured by the quadratic ``verification'' cost function 
\begin{align}
  \Psi(\u) =   \Psi\bigl( \xa_v(\u) \bigr) = \frac{1}{2} \left( \xa_v(\u) - \x^{\rm verif}_v \right)^T \, \C \, \left( \xa_v(\u) - \x^{\rm verif}_v \right) \,. 
  \label{eqn:optobs_psifunc}
\end{align}
The weighting matrix $\C$ can be prescribed to scale the error or to restrict it
to a certain subdomain.
Throughout this paper we take it equal to be the identity matrix, therefore
each component of the error vector $(\xa_v - \x^{\rm verif}_v)$ contributes equally towards the value of $\Psi$.
The analysis $\xa_0$ depends on all sources of information being assimilated: background, observations etc.
The function $\Psi$ depends directly on $\xa_0(\u)$, and therefore depends indirectly on the system parameters $\u$.

The continuous parameter optimization problem seeks the parameter value $\u_\textrm{opt}$ which minimizes the verification functional \eqref{eqn:optobs_psifunc}
\begin{equation}
 \u_\textrm{opt} = \arg\min_{\u} \Psi\bigl(\xa_v\bigr) \quad \textnormal{subject to } 
 \left\{ 
 \begin{array}{l}
 \nabla_{\x_0} \Jfunc\bigl(\xa_0,\u\bigr) = 0\,, \\[2pt]
 \xa_v = \Model_{t_0 \rightarrow t_v} (\xa_0)\,.
 \end{array}
 \right. 
 \label{eqn:optobs_minarg}
\end{equation}
The optimization problem is constrained by the first order optimality condition \eqref{eqn:optobs_fooc},
which implicitly defines the dependency of the analysis $\xa_0(\u)$ on the system parameters $\u$,
and by the model equations, which relate $\xa_v(\u)$ to $\xa_0(\u)$. Note that \eqref{eqn:optobs_minarg}
is an ``optimization-constrained optimization problem'' with the following equivalent formulation 
\begin{equation}
 \u_\textrm{opt} = \arg\min_{\u} \Psi\bigl(\xa_v\bigr) \quad \textnormal{subject to } 
 \left\{ 
 \begin{array}{l}
 \xa_0 = \arg\min_{\x_0} \Jfunc\bigl(\x_0,\u\bigr)\,, \\[2pt]
 \xa_v = \Model_{t_0 \rightarrow t_v} (\xa_0)\,.
\end{array}
 \right. 
 \label{eqn:optobs_minarg_alternative}
\end{equation}

Gradient-based methods can be employed to solve \eqref{eqn:optobs_minarg}.
Each iteration requires computing the value of the cost function $\Psi(\xa_v)$,
which in turn requires solving the 4D-Var optimization problem \eqref{eqn:optobs_fdvar-optimization}   to obtain $\xa_0$.
The two optimization problems \eqref{eqn:optobs_minarg} and \eqref{eqn:optobs_fdvar-optimization} are nested in an outer loop-inner loop fashion.
The inner (4D-Var) optimization \eqref{eqn:optobs_fdvar-optimization} must be solved as accurately as possible 
in order for the optimality condition constraint in \eqref{eqn:optobs_minarg} to hold.
This requires additional computational effort.

Each outer iteration also requires the gradient of $\Psi$ with respect to $\u$.
The first-order optimality condition reads
\begin{align}
 \nabla_\u \Psi\bigl(\xa_v(\u_\textrm{opt})\bigr) = 0\,.
 \label{eqn:optobs_psifooc}
 \end{align}
Using chain-rule differentiation on \eqref{eqn:optobs_psifooc}:
\begin{align}
 \nabla_\u \Psi(\xa_v(\u)) &= \nabla_\u \, \xa_0(\u) \cdot \nabla_{\x_0} \, \xa_v \cdot \nabla_{\xa_v} \Psi(\xa_v) \nonumber \\
                                 &= \nabla_\u \, \xa_0(\u) \cdot \M^T_{0,v} \cdot \C \, \left( \xa_v -\x^{\rm verif}_v \right) \nonumber \\
                              &=    -\nabla_{\u,\x_0}^2 \Jfunc  \cdot \left(\nabla_{\x_0,\x_0}^2 \Jfunc \right) ^{-1}\cdot \M^T_{0,v} \cdot \C \, \left( \xa_v -\x^{\rm verif}_v \right)\,.
 \label{eqn:optobs_psigrad}
\end{align}
Equation \eqref{eqn:optobs_psigrad} is evaluated as follows. The scaled errors at the 
verification time $t_v$ are propagated backwards in time
to $t_0$ via the adjoint model. This vector is then multiplied by the sensitivity of the 
analysis with respect to model parameters \eqref{eqn:dxa-du}. Specifically, a linear system with the 4D-Var Hessian
matrix is solved, and the result multiplied by the second derivative of the 4D-Var cost function. 

The last step in fully defining the metaoptimization problem for the data assimilation system is to specify the system parameters $\u$ that we seek to optimize.
In this paper we consider three particular 4D-Var parameters:
\begin{enumerate}
 \item Observation values: $\u \equiv \y$,
 \item Observation error covariance: $\u \equiv \R_k^{-1}$, and
 \item Observation locations (in 2D space): $\u \equiv (\ell_x,\ell_y)$.
\end{enumerate}

Each of the three choices for $\u$ is defined in a continuous space.
Measurements can take any value within the allowable range of the observed physical variable.
Error covariances are defined from the expected value and standard deviation.
Sensor locations represent physical locations in the space domain.
Neither one imposes any restrictions in defining the continuous optimization problem.

It is important to note that the verification cost function \eqref{eqn:optobs_psifunc} is not the only metric of the data assimilation system 
one can seek to improve.
Previous work in experimental design to optimize sensor networks for direct state estimation 
(unconstrained by a PDE system) defines the verification cost function based on the Fisher Information Matrix
corresponding to the observations \cite{ucinski2000optimal}. The objective is to optimize
an algebraic aspect of the information matrix, related to its spectrum or determinant.
Alternatively, an optimal placement of sensors for data assimilation has been proposed
such as to maximize the dynamical system observability \cite{TELLUSA17133}.

\subsection{Optimization of observation values}\label{sec:optimize-values}

The first application for our continuous optimization approach is reconstructing
the optimal values of assimilated observations, i.e observations that would lead
to a 4D-Var analysis as close as possible to the verification $\xv_0$.
The mismatches 
between the original dataset and the optimal dataset (\textit{``what we should have measured''}),
are useful to detect faulty sensors or errors in the handling and processing of measurement data.

This is achieved by minimizing the verification cost function $\Psi$ with respect 
to the values of selected observations:
\begin{equation}
 \label{eqn:optimal-y}
  (\y_k)_\textrm{opt} = \arg\min_{\y_k} \Psi\bigl(\xa_v\bigr) \quad \textnormal{subject to }
 \left\{ 
 \begin{array}{l}
 \nabla_{\x_0} \Jfunc\bigl(\xa_0,\y_k\bigr) = 0\,, \\[2pt]
 \xa_v = \Model_{t_0 \rightarrow t_v} (\xa_0)\,.
 \end{array}
 \right. 
\end{equation}
A numerical solution requires the gradient of $\Psi$ with respect to observations.
Differentiating the optimality condition \eqref{eqn:optops_fdvar-fooc} with respect to $\y_k$ gives
\[
\nabla_{\y_k, \x_0}^2 \, \Jfunc(\xa_0) = -\R_k^{-1} \, \HH_k \, \M_{0,k} \,,
\]
which is used in \eqref{eqn:optobs_psigrad} to obtain the following expression for the gradient of the verification functional to observations:
\[
 \nabla_{\y_k} \Psi = \R_k^{-1}\, \HH_k\, \M_{0,k}\, \left(\nabla_{\x_0,\x_0}^2 \Jfunc(\xa_0)\right) ^{-1} \, \M^T_{0,v} \, \C \, \left( \xa_v -\x^{\rm verif}_v \right)\,.
\]

\subsection{Optimization of observation weights}\label{sec:optimize-weights}

The second application of our framework is the tuning of observation covariances.
Each individual data point $(\y_k)_{(i)}$ has an associated error variance $(\sigma^2_k)_{(i)}$, 
a scalar value which serves as a weighting coefficient during assimilation.
The values of $(\sigma^2_k)_{(i)}$ reflect the level of trust one has in the data.
This can be related to sensor accuracy and other characteristics of the measurement process.

We use our framework to obtain the optimal values for the $(\sigma^2_k)_{(i)}$ weights 
for which the verification cost function $\Psi$ is minimized.
The optimized weights are obtained through a dynamic approach that will reflect 
how the data assimilation process made use of the input data (observations).
Smaller standard deviations will be associated to observations which have a significant contribution,
while larger standard deviations will reflect redundant or corrupt data.
It is expected that the data assimilation problem formulated with the improved weights will provide superior forecasts.
In addition, the optimal weight values can be used
to configure the sensor scanning strategy in terms of intensity or frequency,
or to remove certain measuring instruments from the network.

In the derivation below we assume that observation error covariances at each time instant are 
diagonal matrices $\R_k = \textnormal{diag}\{(\sigma^2_k)_{(i)}\}$.
The optimization variables in our formulation are the diagonal elements of all $\R_k$, and 
we seek to compute the optimal values $(\sigma^2_k)_{(i)}$ for which $\Psi$ attains its minimum:
\begin{equation}
 \label{eqn:optimal-R}
 (\R_k)_\textrm{opt} = \arg\min_{\R_k} \Psi\bigl(\xa_v\bigr) \quad \textnormal{subject to }
 \left\{ 
 \begin{array}{l}
 \nabla_{\x_0} \Jfunc\bigl(\xa_0,\R_k\bigr) = 0\,, \\[2pt]
 \xa_v = \Model_{t_0 \rightarrow t_v} (\xa_0)\,.
 \end{array}
 \right. 
\end{equation}
From the 4D-Var sensitivity equations \cite{Daescu_2008} the gradient of $\Psi$ to $\R_k$ is:
\begin{align}
 \nabla_{\R_k} \Psi = \left( \R_k^{-1} \left[ \H(\x) - \y \right] \right) \, \otimes \, \nabla_\y \Psi(\R_k)\,,
 \label{eqn:optobs_senscov}
\end{align}
where the operator $\otimes$ denotes the Kronecker matrix product, and 
 the expression \eqref{eqn:optobs_senscov} holds for non-diagonal $\R_k$ matrices as well.

Previous studies  \cite{Daescu_2010} have performed background and error covariance tuning by 
parameterizing the error covariance matrices as
\[
 \B_0(s_0) = s_0 \cdot \B_0, \qquad \R_k(s_k) = s_k \cdot \R_k\,,
\]
and by optimizing the scaling parameters $s_0$ and $s_k$ such that more weight
is associated either to the background  or to the observations.

Our approach differentiates between individual observation weights
and optimizes for each component of $\R_k$. Based on \eqref{eqn:optobs_senscov}
the optimization problem \eqref{eqn:optimal-R} can be formulated and solved for
non diagonal observation error covariances as well.

\subsection{Optimization of observation locations}\label{sec:optimize-locations}

The third application of our optimization framework is finding the optimal locations for the sensors.

In a real data assimilation system observations are available at a sparse set of preselect locations where the measuring instruments are installed.
We consider the case where the measured quantities correspond to computed variables in the model.
The observation operator $\H$ \eqref{eqn:optobs_fdvar-costfun} interpolates the model solution $\x_k$ in physical space to the locations of 
the sensors measuring $\y_k$.
Each component of the observation vector $\y_k$ is associated with its location in physical space, e.g., in two dimensions
identified by the Cartesian coordinates $(\ell_x,\ell_y)$.
Similarly, each component  $\z_i$ of the model solution $\z \equiv \x_k$ is associated with a grid point with coordinates  $(\ell_{x_i},\ell_{y_i})$. 

The formulation of the interpolation operator $\H_k$ explicitly contains the coordinates of the sensor locations
(i.e. of the points where interpolation is performed). Therefore the 4D-Var cost function $\Jfunc$, and the analysis $\xa_0$, both depend on these locations.
We denote the set of all sensor locations within the assimilation window by 
\[
L = \bigcup_{\rm all~sensors} \{(\ell_x,\ell_y,t_k)\}\,.
\]
The sensor location optimization problem is formulated as follows: 
\begin{align}
 \label{eqn:optimal-loc}
 L_\textrm{opt} &= \arg\min_{\ell_x,\ell_y} \Psi\bigl(L) \quad \textnormal{subject to }
 \left\{ 
 \begin{array}{l}
 \nabla_{\x_0} \Jfunc\bigl(\xa_0,L\bigr) = 0\,, \\[2pt]
 \xa_v = \Model_{t_0 \rightarrow t_v} (\xa_0)\,.
 \end{array}
 \right. 
\end{align}
In this study we choose the Inverse Distance Weighting (IDW) \cite{shepard1968two} interpolant as the $\H_k$ operator.
We apply it to the spatial locations only and do not attempt to optimize for the time distribution of measurements.
IDW is a general scheme and is very popular in geographic information systems and climate modeling. 

IDW interpolates the model solution values $\z_i$ through a linear combination 
of all available data points, each weighted by their inverse distance to the interpolation location.
For each interpolation location $(\ell_x, \ell_y)$ we have
\begin{eqnarray}
 &&\H_k(\ell_x, \ell_y; \z) = \left\{
 \begin{array}{cl}
 \displaystyle
 \frac{\sum\limits_{i} d_i^{-1} \, \z_i}{\sum\limits_{i}d_i^{-1} },& \textnormal{if } d_i \ne 0\,,  \label{eqn:optobs_idw} \\ 
  \z_i,& \textnormal{if } d_i = 0\,,
 \end{array}
 \right.\\ 
&& \textnormal{where    } d_i = \left[ \left(\ell_x - \ell_{x_i}\right)^2 + \left( \ell_y - \ell_{y_i} \right)^2 \right]^{1/2} \nonumber\,.
\end{eqnarray}

Differentiation of the first order optimality condition of 4D-Var \eqref{eqn:optops_fdvar-fooc} yields
%
%
\begin{subequations}
\label{eqn:dJdL}
\begin{eqnarray}
\label{eqn:dJdLx}
\nabla_{\ell_x,\x_0}^2 \Jfunc &=& \nabla_{\ell_x} \H_k(\ell_x,\ell_y; \xa_k) \\
\nonumber
&& +  \M_{0,k}^T \, \left(\nabla_{\ell_x} \HH_k(\ell_x,\ell_y; \xa_k)\right)^T \, \R_k^{-1}\, \left( \Hobs_k(\ell_x,\ell_y; \xa_k) - \y_k \right)  \nonumber \,, \\
\label{eqn:dJdLy}
\nabla_{\ell_y,\x_0}^2 \Jfunc &=& \nabla_{\ell_y} \H_k(\ell_x,\ell_y; \xa_k) \\
\nonumber
&& +  \M_{0,k}^T \, \left(\nabla_{\ell_y} \HH_k(\ell_x,\ell_y; \xa_k)\right)^T \, \R_k^{-1} \left( \Hobs_k(\ell_x,\ell_y; \xa_k) - \y_k \right)  \nonumber \,.
\end{eqnarray}
\end{subequations}
The gradient of $\Psi$ to coordinates $\ell_x$ and $\ell_y$ is then computed 
from \eqref{eqn:optobs_psigrad}
\begin{eqnarray}
 \label{eqn:optobs_psigradh}
 \nabla_{\ell_x} \Psi = \nabla_{\ell_x,\x_0}^2 \Jfunc \cdot \left(\nabla_{\x_0,\x_0}^2 \Jfunc(\xa_0)\right) ^{-1} \, \M^T_{0,v} \, \C \, \left( \xa_v -\x^{\rm verif}_v \right)\,, \\
\nonumber
 \nabla_{\ell_y} \Psi =\nabla_{\ell_y,\x_0}^2 \Jfunc \cdot \left(\nabla_{\x_0,\x_0}^2 \Jfunc(\xa_0)\right) ^{-1} \, \M^T_{0,v} \, \C \, \left( \xa_v -\x^{\rm verif}_v \right)\,.
\end{eqnarray}

Differentiation of \eqref{eqn:optobs_idw} with respect to $\ell_x$, $\ell_y$ gives the gradients of $\H_k$ with respect to sensor locations:
\[
 \nabla_{\ell_x} \H_k(\ell_x, \ell_y; \z) = \frac{\sum\limits_i \sum\limits_j (d_i)^{-2} (d_j)^{-1} \left(\ell_x - \ell_{x_i}\right) \z_i \left( \z_i - \z_j \right)}{ \left| \sum\limits_i (d_i)^{-1} \right|^2 }\,,
\]
and
\[
 \nabla_{\ell_y} \H_k(\ell_x, \ell_y; \z) = \frac{\sum\limits_i \sum\limits_j (d_i)^{-2} (d_j)^{-1} \left(\ell_y - \ell_{y_i}\right) \z_i \left( \z_i - \z_j \right)}{ \left| \sum\limits_i (d_i)^{-1} \right|^2 }\,.
\]
The innovation vectors $\Hobs_k(\ell_x,\ell_y; \xa_k) - \y_k $ evaluated at the analysis solution are likely to have small values.
For this reason the derivatives \eqref{eqn:dJdL} are well approximated by the simpler expressions
\begin{equation}
\label{eqn:dJdL-approximate}
\nabla_{\ell_x,\x_0}^2 \Jfunc \approx \nabla_{\ell_x} \H_k(\ell_x,\ell_y; \xa_k)\,, \quad
\nabla_{\ell_y,\x_0}^2 \Jfunc \approx \nabla_{\ell_y} \H_k(\ell_x,\ell_y; \xa_k)\,, 
\end{equation}
which do not require additional adjoint integrations.

\section{Applications}\label{sec:optobs_appl}

We illustrate the proposed metaoptimization approach for 
the three applications presented in Section \ref{sec:optobs_contopt} using the
two dimensional shallow water equations system.

\subsection{Test problem}

The two-dimensional shallow-water equations (2D {\sc swe})  \cite{CPA:CPA3160210103} approximate the movement of a thin layer of fluid inside a basin:
\begin{eqnarray}
 \frac{\partial}{\partial t} h + \frac{\partial}{\partial x} (uh) + \frac{\partial}{\partial y} (vh) &=& 0 \nonumber \\
 \frac{\partial}{\partial t} (uh) + \frac{\partial}{\partial x} \left(u^2 h + \frac{1}{2} g h^2\right) + \frac{\partial}{\partial y} (u v h) &=& 0  \label{swe} \\
 \frac{\partial}{\partial t} (vh) + \frac{\partial}{\partial x} (u v h) + \frac{\partial}{\partial y} \left(v^2 h + \frac{1}{2} g h^2\right) &=& 0 \;.
\nonumber
\end{eqnarray}
Here $h(t,x,y)$ is the fluid layer thickness, and $u(t,x,y)$ and $v(t,x,y)$ are the components of 
the velocity field of the fluid. The gravitational acceleration is denoted by $g$. 

We consider on a spatial domain $\Omega = [-3,\,3]^2$ (spatial units), 
and an integration window is $t_0 = 0 \le t \le t_\textrm{f} = 0.1$ (time units). 
Boundary conditions are specified periodic. The space discretization is realized using a finite volume scheme, 
and the time integration uses a fourth Runge-Kutta scheme, following the method Lax-Wendroff \cite{Wendroff_1998}.
The model uses a square $q \times q$ uniform spatial discretization grid, which brings the number of model (state) variables to $n = 3\,q^2$.

Our adjoint models are built using the automatic differentiation tool TAMC \cite{giering1997tangent,TAMC}. 
Perturbations are propagated forward in time using the tangent-linear model ({\sc tlm}),
and backwards in time using the first-order adjoint model ({\sc foa}), which can also efficiently computes 
the gradient of a cost functional defined on the model states. 
The product between the Hessian of a cost functional and a user-defined vector can be computed with 
the second-order adjoint model ({\sc soa}) \cite{Cioaca_2011}. 

%
%

\subsection{Data assimilation setting}\label{sec:optobs_dasetup}

The 4D-Var system is set up for a simple version of the ``\textit{circular dam}'' problem \cite{anastasiou1997solution}.
Where not specified otherwise, the 4D-Var parameters are configured as follows.

\begin{figure}
\centering
\subfigure[Initial time]{ \includegraphics[width=0.45\textwidth]{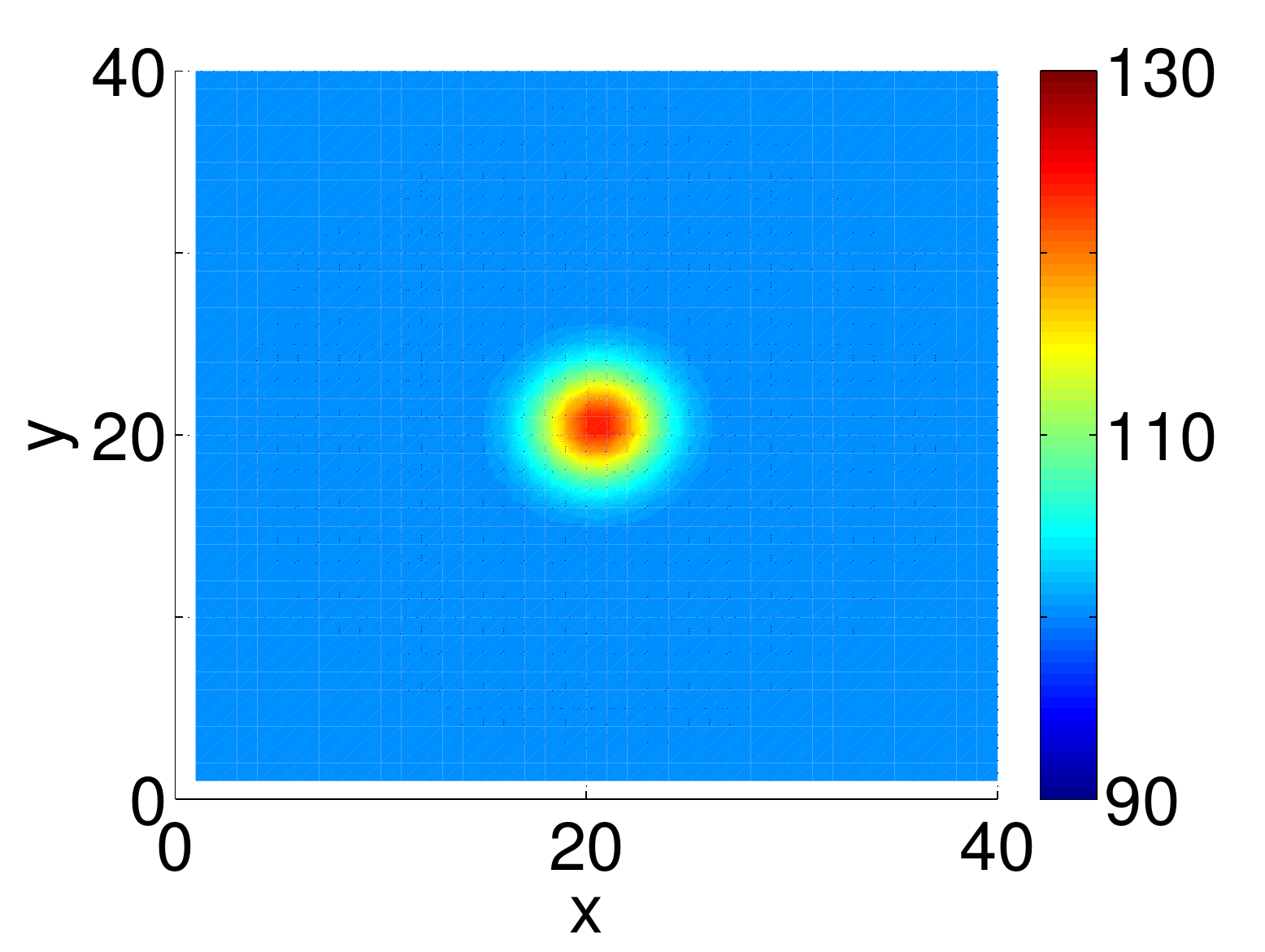} }
\subfigure[Final time]{ \includegraphics[width=0.45\textwidth]{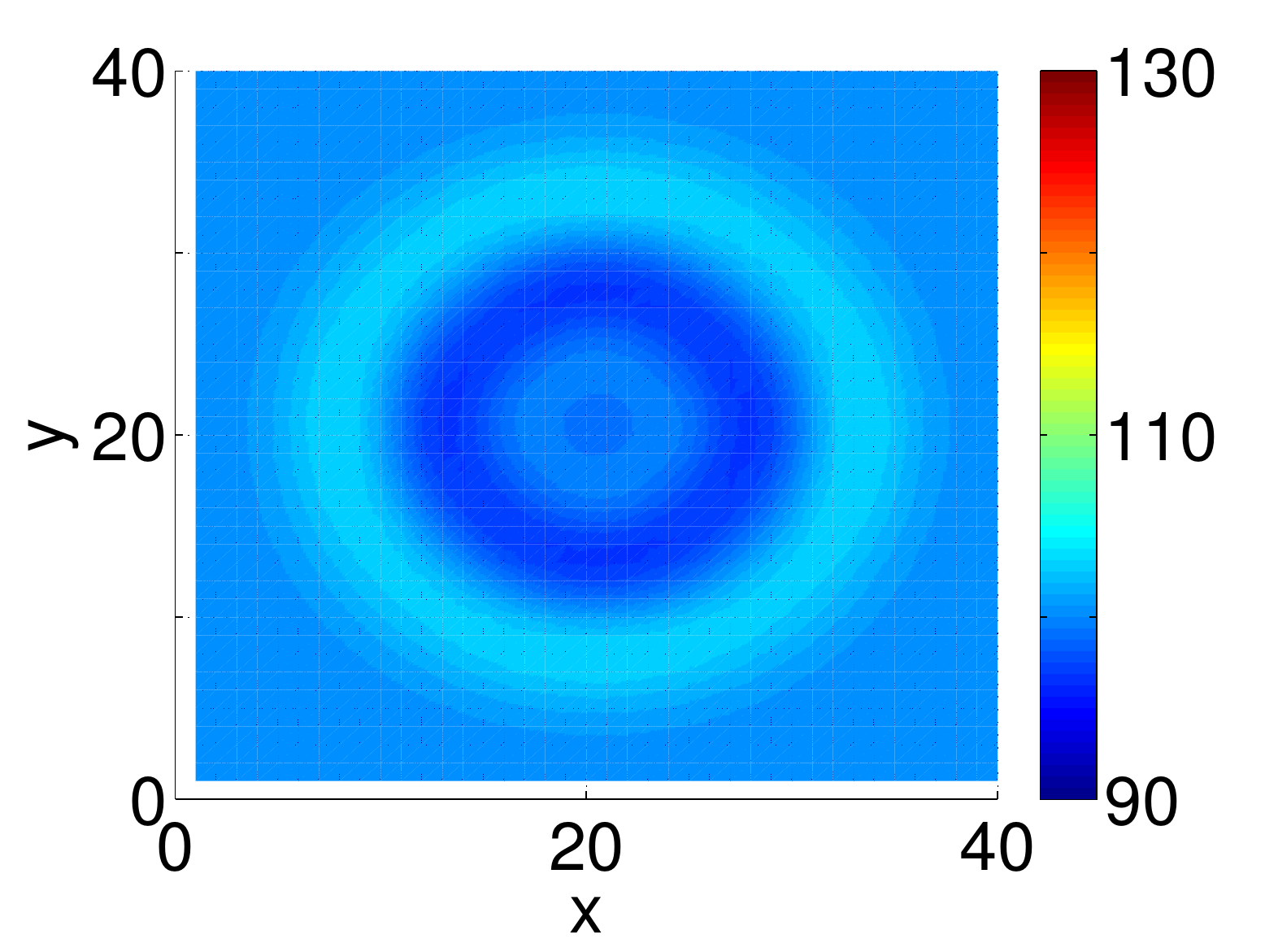} }
\caption{The height field $h$ at the beginning and at the end of the reference trajectory.}
\label{fig:optobs_swetraj}
\end{figure}

The reference initial height field $h$ is a Gaussian bell of a width equal to $10$ gridpoints centered at the midpoint of the grid, 
and the reference initial velocity vector components are constant $u=v=0$. 
The physical interpretation is that the front of water falls to the ground ($h$ decreases) under the effect of gravity
and creates concentric ripples which propagate towards the boundaries. 
Figures \ref{fig:optobs_swetraj} represent snapshots of the reference trajectory at initial and final time. 

The computational grid is square and regular with $q=40$ grid points in each direction, for a total of $4800$ model variables (states).
The simulation time interval is set to $0.01$ seconds, using $N=100$ timesteps of size $0.0001$ (time units).

The $h$ component of the a priori estimate (background) $\xb_0$ is generated by adding a correlated perturbation to the $h$ reference solution at initial time.
The background error covariance $\B_0$ corresponds to a standard deviation of $5\%$ of the reference field values.
For the $u$ and $v$ components we use white noise to prescribe perturbations.
The spatial error correlation uses a Gaussian decay model, with a correlation distance of five. 

Synthetic observations are generated at the final time $t_{100}$, by adding random noise to the reference trajectory.
Since the observation errors are assumed uncorrelated, the observation error covariance matrix $\R_{100}$ is diagonal.
The standard deviation for observation noise is $1\%$ of the largest absolute value of the observations for each variable.
We consider observations of all variables at each grid point and the observation operator $\H$ to be linear. 
The minimization of the 4D-Var cost function is performed with the 
L-BFGS-B solver \cite{zhu1997algorithm} using a fixed number of $100$ iterations.
For our problem, this guarantees that the analysis is sufficiently close to the minimum.

For the verification cost function $\Psi$ we choose $\C$ to be the identity matrix and the verification time $t_v = t_{100}$,
which corresponds to the final time of our assimilation window.
The meta-optimization problems \eqref{eqn:optimal-y}, \eqref{eqn:optimal-R}, and \eqref{eqn:optimal-loc} are also solved using
 L-BFGS-B \cite{zhu1997algorithm}, which performs 10 ``outer'' iterations to minimize $\Psi(\u)$ with respect to $\u$. Each outer iteration requires
 the solution of a 4D-Var data assimilation problem, which is obtained by running 100 ``inner iterations''
 of L-BFGS-B to minimize $\mathcal{J}(\x_0,\u)$ with respect to $\x_0$.


\subsection{Optimization of observation values}

This experiment employs the methodology developed in Section \ref{sec:optimize-values} to correct the
values of faulty observations; this is made possible by the redundant information available in the verification functional.

\begin{figure}
\setcounter{subfigure}{0}
\centering
 \subfigure[Initial Time]{ \includegraphics[width=0.45\textwidth]{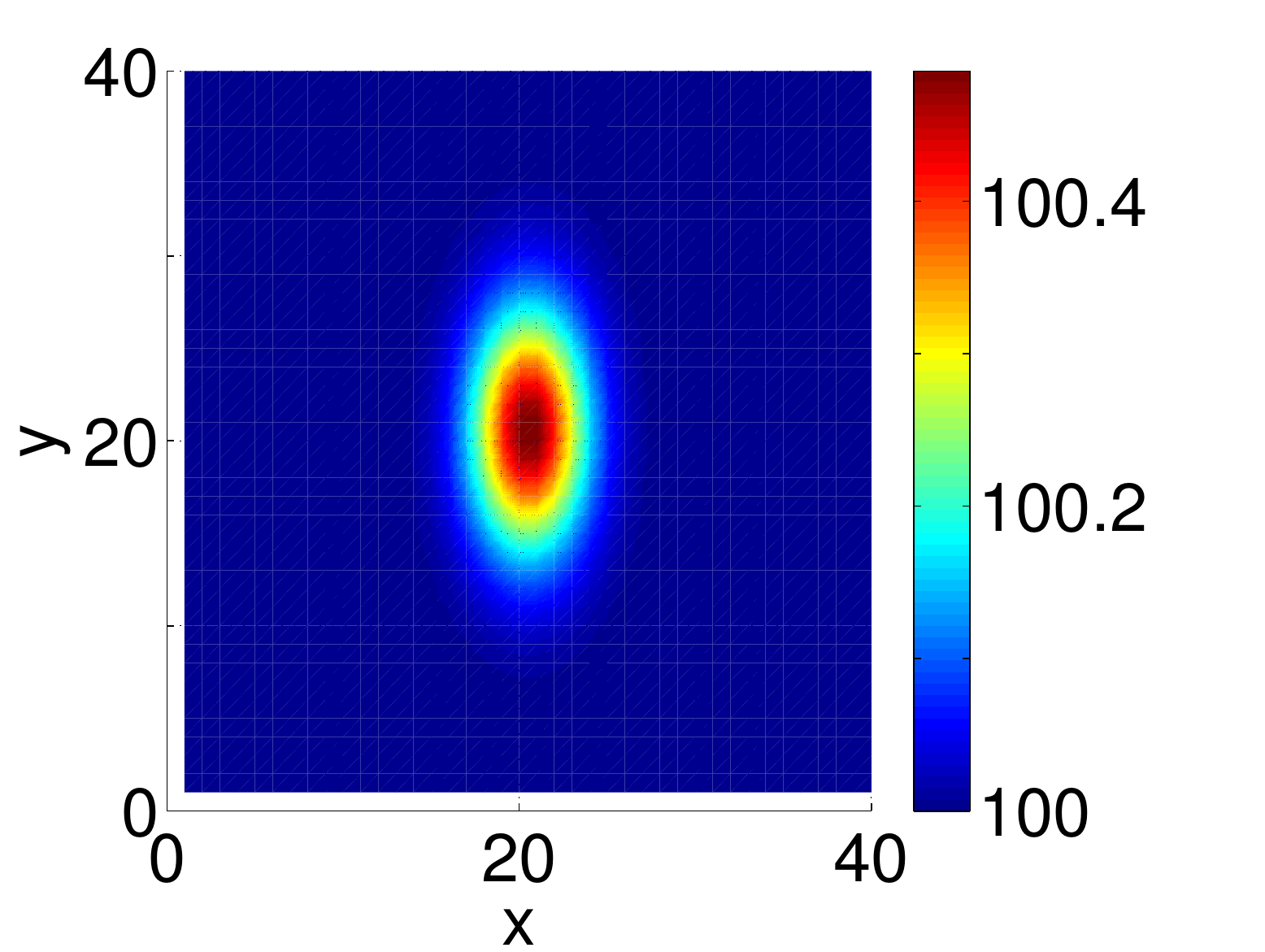} \label{fig:optobs_OBS_init_NS} }
 \subfigure[Observation Time]{ \includegraphics[width=0.45\textwidth]{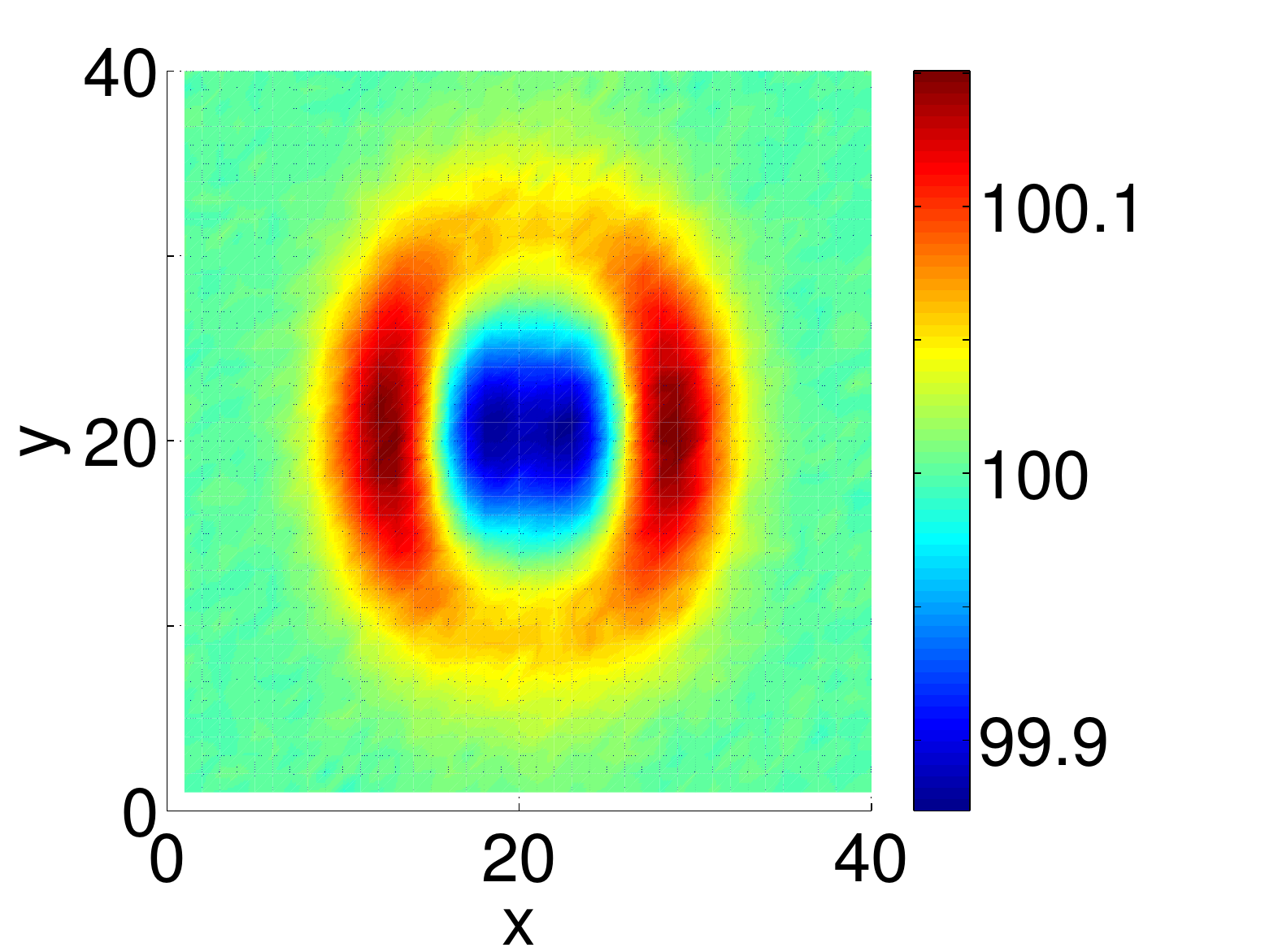} \label{fig:optobs_OBS_refobs} } \\
 \caption{The reference height field $h$ at the initial and final (observation) times.}
\end{figure}

The reference height field $h$ at the initial time is shown in Figure \ref{fig:optobs_OBS_init_NS},
and reference solution at the final time $t_{100}$  in Figure \ref{fig:optobs_OBS_refobs}; the latter is
used to generate synthetic observations.
Assimilation of perfect observations in a twin experiment framework 
yields an analysis that is close to the reference.

\begin{figure}
\setcounter{subfigure}{0} 
\centering
 \subfigure[Faulty Observations]{ \includegraphics[width=0.45\textwidth]{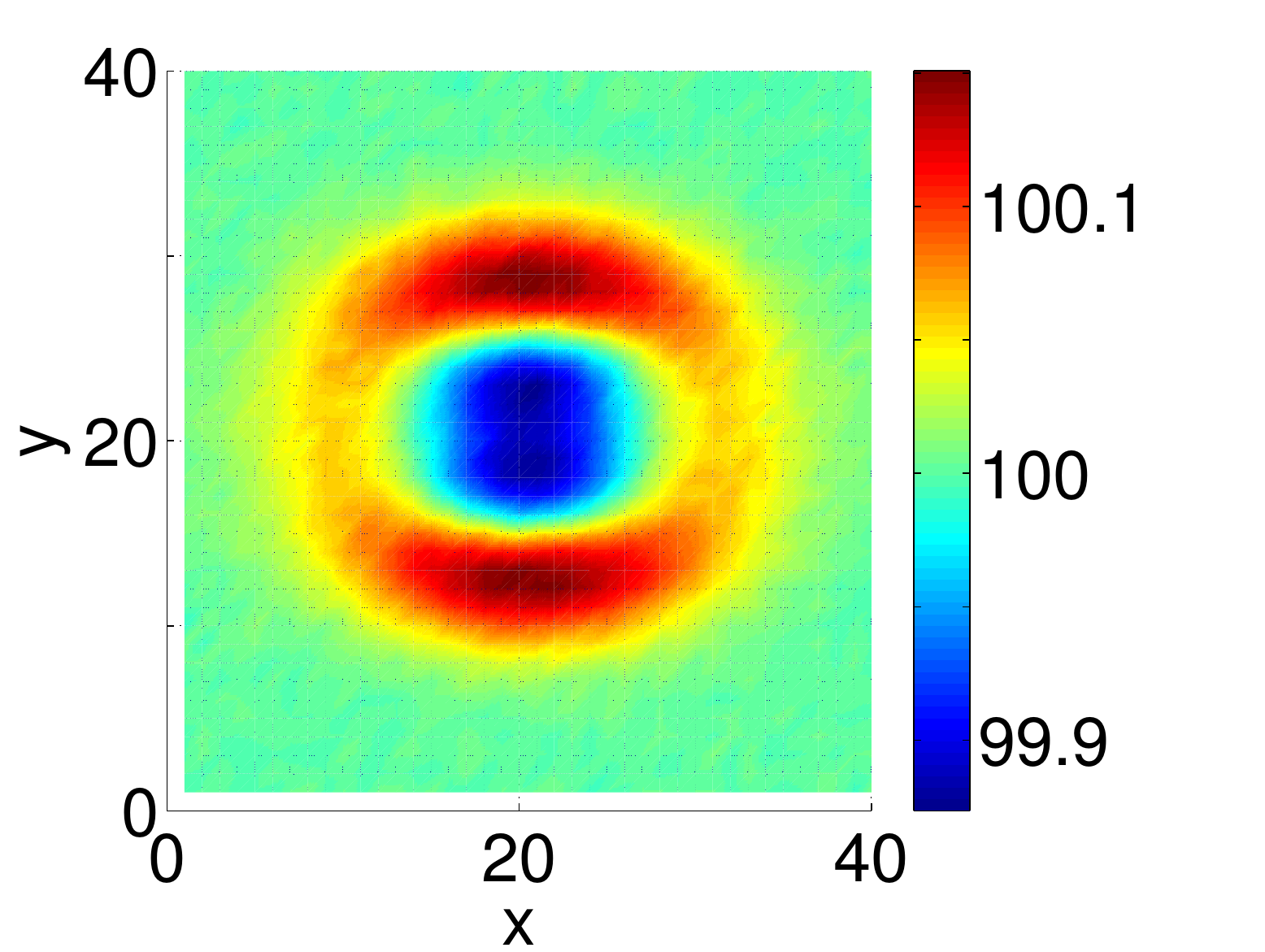} \label{fig:optobs_OBS_initobs} } 
 \subfigure[4D-Var Analysis]{ \includegraphics[width=0.45\textwidth]{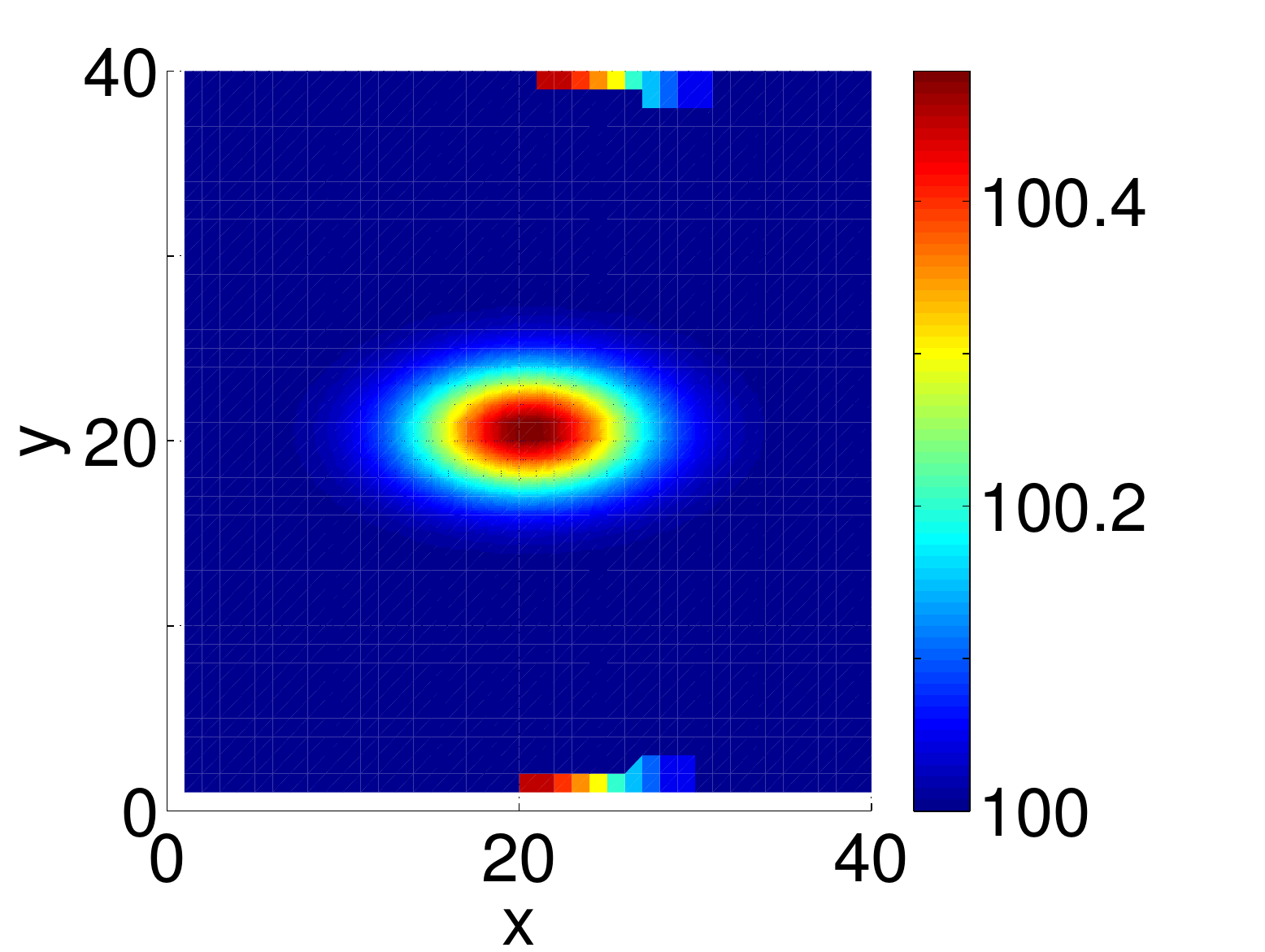} \label{fig:optobs_init_EW} } \\
\caption{The faulty (unoptimized) observations of the height field $h$ and the corresponding 4D-Var analysis.}
\end{figure}

Here we consider the case where some of the observations 
are corrupted. Specifically, we consider synthetic observations corresponding to the field
shown in Figure \ref{fig:optobs_OBS_initobs}. The difference between these corrupted
observations and the reference ones (Figure \ref{fig:optobs_OBS_refobs})
 consists in the orientation of the field, now aligned along the perpendicular axis. 
For example, this could be the result of an indexing mistake when processing large data.
Then assimilation of this faulty dataset yields the $h$ analysis shown in Figure \ref{fig:optobs_init_EW}, 
and is clearly distinct from the reference solution (Figure \ref{fig:optobs_OBS_init_NS}).

We apply the methodology developed in Section \ref{sec:optimize-values} to define and solve an 
optimization problem that corrects the observation values $\y_k$ such that the 4D-Var analysis becomes 
more accurate (i.e., closer to the reference).

\begin{figure}
\setcounter{subfigure}{0}
\centering
 \subfigure[L-BFGS convergence]{ \includegraphics[width=0.45\textwidth]{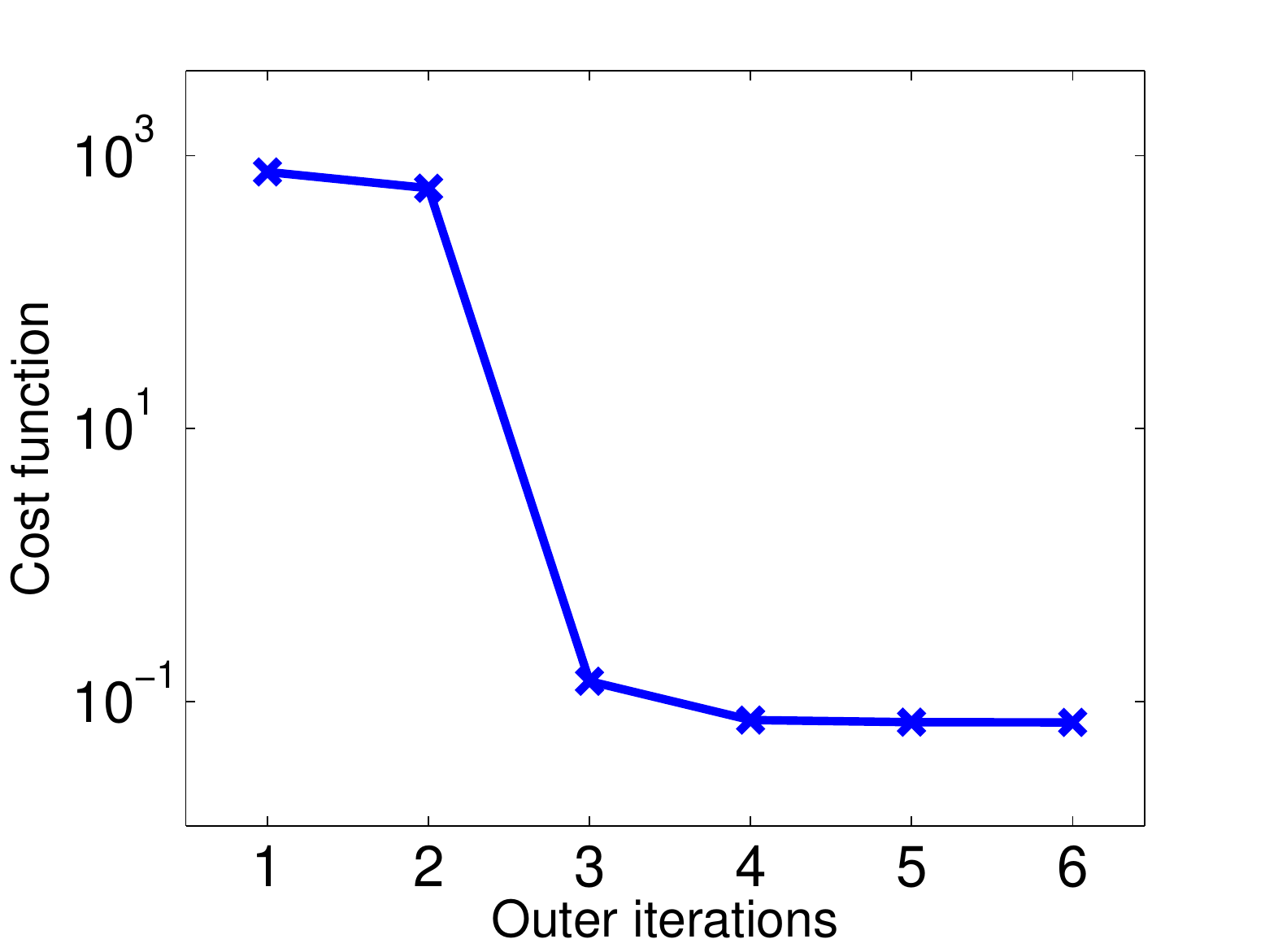} \label{fig:optobs_OBS_cost} }
 \subfigure[Optimized observation values]{ \includegraphics[width=0.45\textwidth]{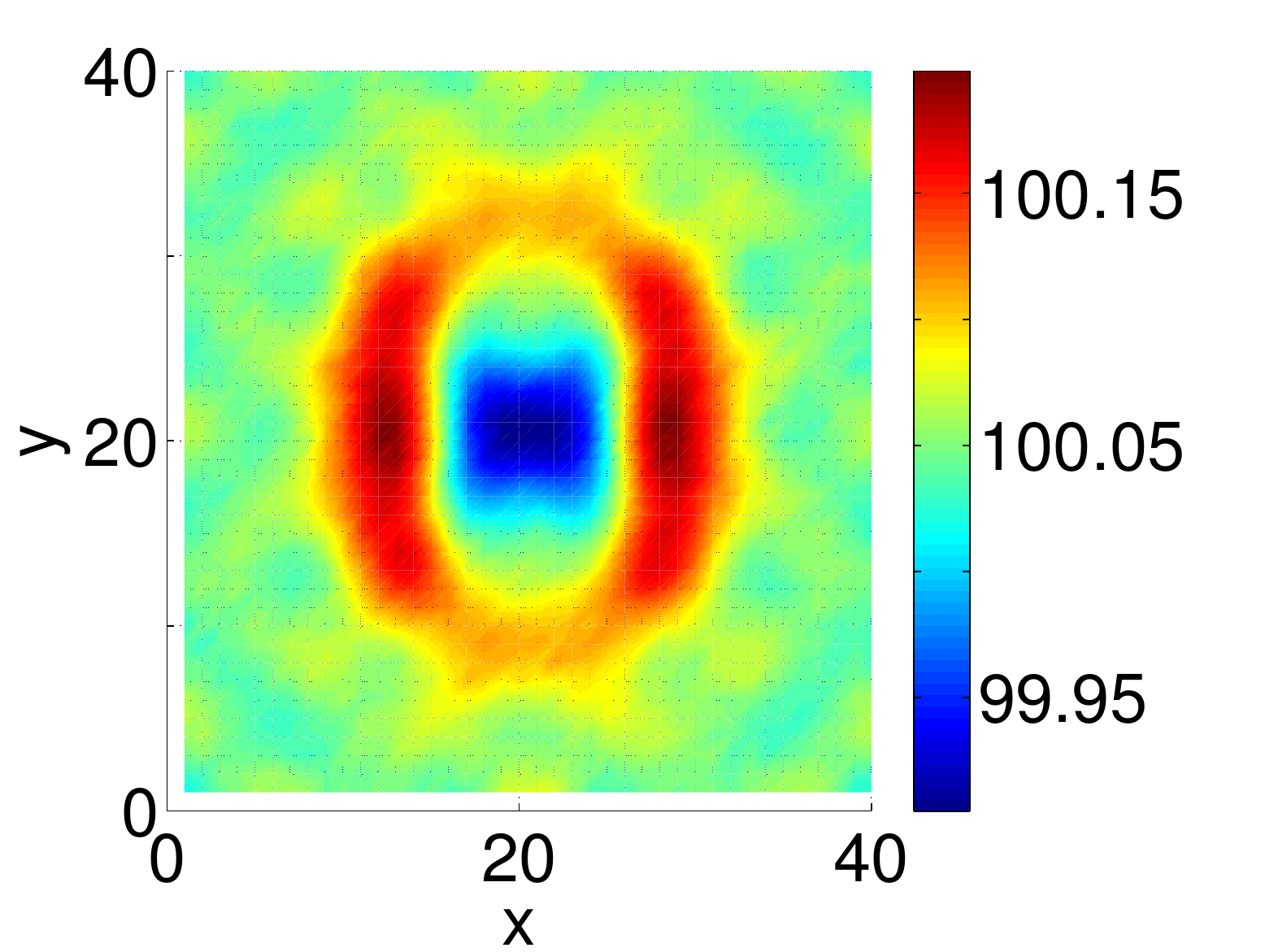} \label{fig:optobs_OBS_optobs} }
 \caption{The minimization of the verification cost function \eqref{eqn:optimal-y} and the optimized observations at assimilation time $t_{100}$.}
\label{fig:optobs_OBSVAL}
\end{figure}

The initial guess for $\y_k$ are the faulty observations from Figure \ref{fig:optobs_OBS_initobs}.
We solve the optimization problem \eqref{eqn:optimal-y} using five iterations of L-BFGS \cite{zhu1997algorithm} 
and plot the decrease of the cost function values $\Psi(\y_k)$ at each iteration in Figure \ref{fig:optobs_OBS_cost},
At first, the error between the 4D-Var analysis and the verification is large,
but then it decreases monotonically with each iteration.
Most of the cost function decrease happens at the second iteration,
with subsequent iterations contributing less significantly to the convergence.
The optimal observations computed after five iterations are plotted in Figure \ref{fig:optobs_OBS_optobs}.
They are similar to the reference values in Figure \ref{fig:optobs_OBS_refobs}. This validates the approach developed
in Section \ref{sec:optimize-values}. Note that the process is computationally efficient, with the optimization
converging in less than five iterations.
The true error norm corresponding to the reanalyzed initial solution $\xa_0$ before and after the optimization 
is shown in Table \ref{Table:obsopt_gain1} and confirms that the 4D-Var process benefits from the new observation values.

\begin{table}
\caption{Error norm of the 4D-Var reanalyzed initial solution before and after optimizing observation values.}
\centering
{
\begin{tabular}{|c||c|c|}
  \hline
 &  Initial observations & Optimized observations \\
 \hline\hline
$\Vert \xa_0 - \x_0^{\rm reference} \Vert$ &  $1.8216$ & $0.0703$\\
 \hline
\end{tabular}
}
\label{Table:obsopt_gain1}
\end{table}

\subsection{Optimization of observation weights}

This experiment illustrates the methodology developed in Section \ref{sec:optimize-weights}.
In practice measurements contain various amounts of noise,
for example due to various sensor accuracies.
The observation error covariances specified initially are only rough approximations of
the true error statistics. We seek to tune the observation error covariances by solving the optimization 
problem \eqref{eqn:optimal-R}. Improved values of observation error statistics can be obtained 
by taking advantage of the additional information encapsulated by the verification cost function.
 
\begin{figure}
\setcounter{subfigure}{0}
\centering
 \subfigure[Perfect Observations]{ \includegraphics[width=0.3\textwidth]{reftraj100_lg} \label{fig:optobs_VAR_obspert} }
 \subfigure[Prescribed Noise (physical variable units)]{ \includegraphics[width=0.3\textwidth]{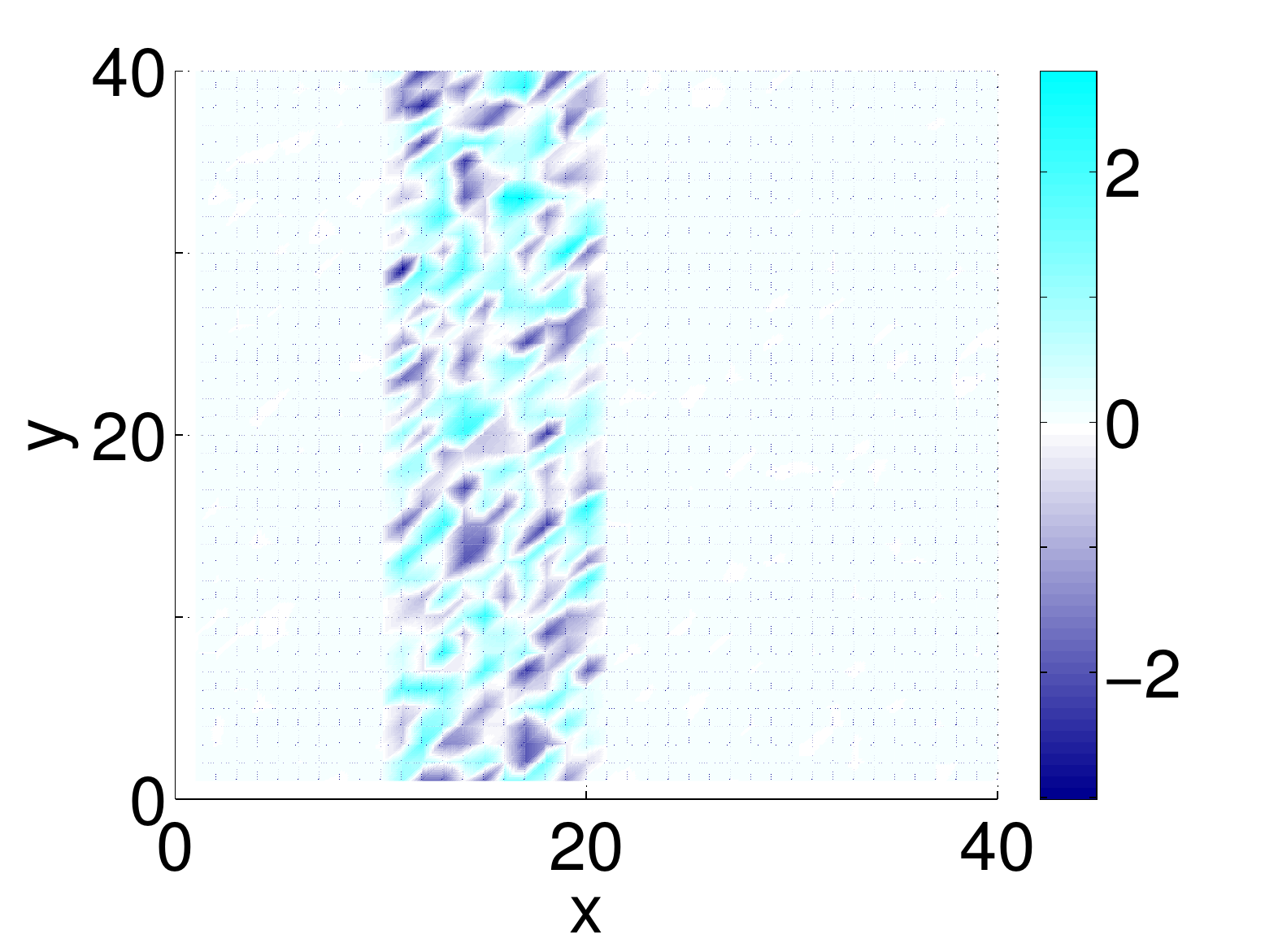} \label{fig:optobs_VAR_obsnoise} }
 \subfigure[4D-Var Analysis]{ \includegraphics[width=0.3\textwidth]{reftraj0_lg} \label{fig:optobs_VAR_init} } \\
 \caption{The $h$ observations, the prescribed observation noise, and the resulting 4D-Var analysis using the initial specification of the error covariances.}
\end{figure}

The experiment is set up as follows. We construct a synthetic observational data set for $h$
from the model trajectory at the final time (Figure \ref{fig:optobs_VAR_obspert}).
White noise is added to the perfect values to simulate observational errors. 
The noise magnitude is nonuniform: it is larger in a contiguous area of rectangular shape,
as illustrated in Figure \ref{fig:optobs_VAR_obsnoise}. Initially the data assimilation system uses the observation error 
covariance $\R_k = \mathbf{I}$. The same noise levels are specified for all grid points, and the data assimilation system is initially unaware of the
larger observation errors in the selected area.

Data assimilation using the initially specified error covariances yields the analysis shown in Figure \ref{fig:optobs_VAR_init}.
This analysis is similar to the reference initial solution (Figure \ref{fig:optobs_swetraj}), and is not indicative
of a misspecification of observation error covariances.
The small mismatch, however,  can be quantified numerically through the verification functional $\Psi$.

The initial observation variances (diagonal elements of $\R_k$) are equal to $1$ (unit), 
meaning that all data points are equally trusted. We now seek to improve the values of the observation error covariances, and thus improve the performance of
the data assimilate system. This is done by solving the optimization 
problem \eqref{eqn:optimal-R} for the diagonal entries of $\R_k$, when the data values $\y$ are fixed.

\begin{figure}
\setcounter{subfigure}{0}
\centering
 \subfigure[L-BFGS Convergence]{ \includegraphics[width=0.3\textwidth,height=0.25\textwidth]{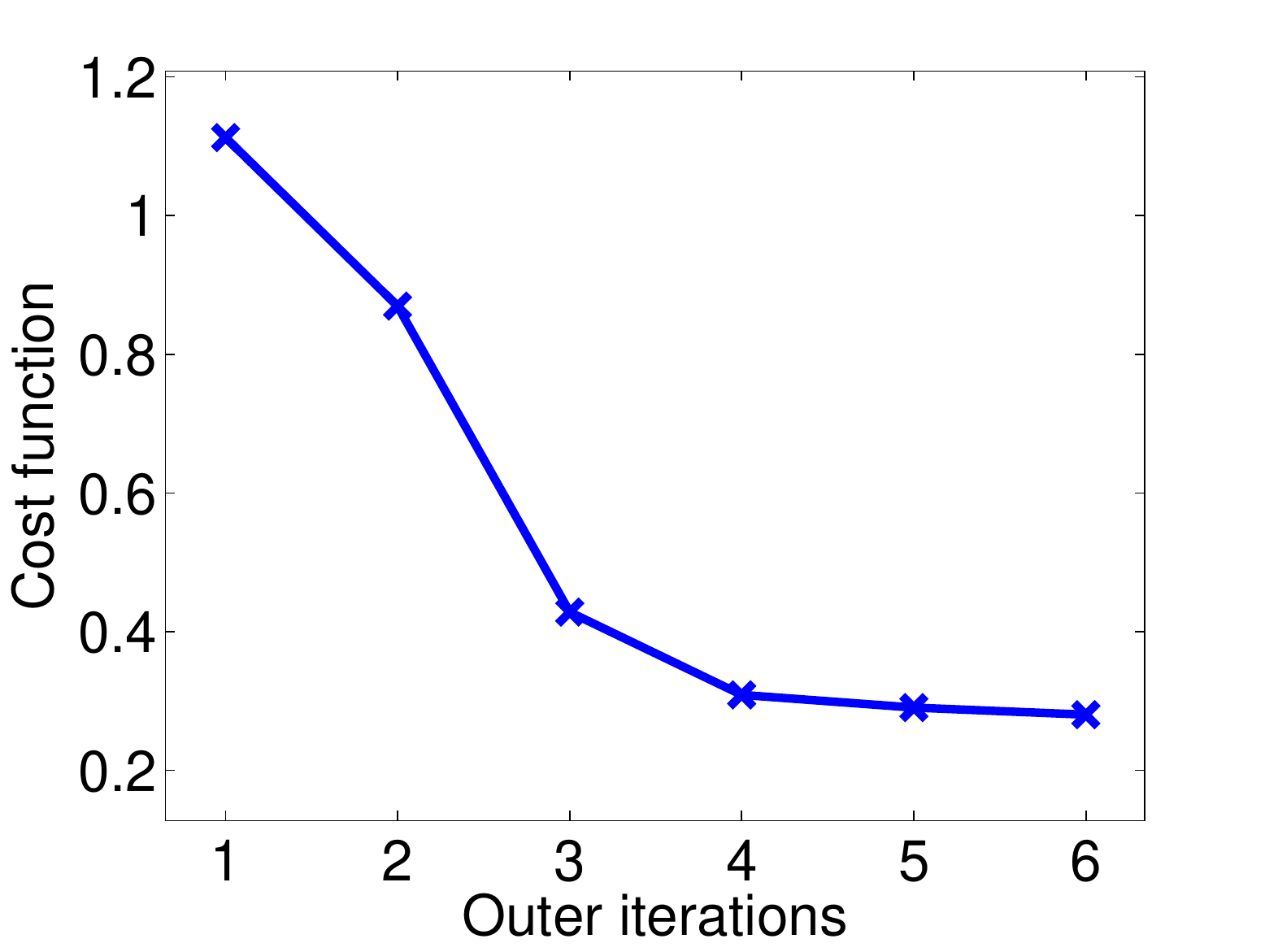} \label{fig:optobs_VAR_cost} }
 \subfigure[Optimized Covariances]{ \includegraphics[width=0.31\textwidth,height=0.25\textwidth]{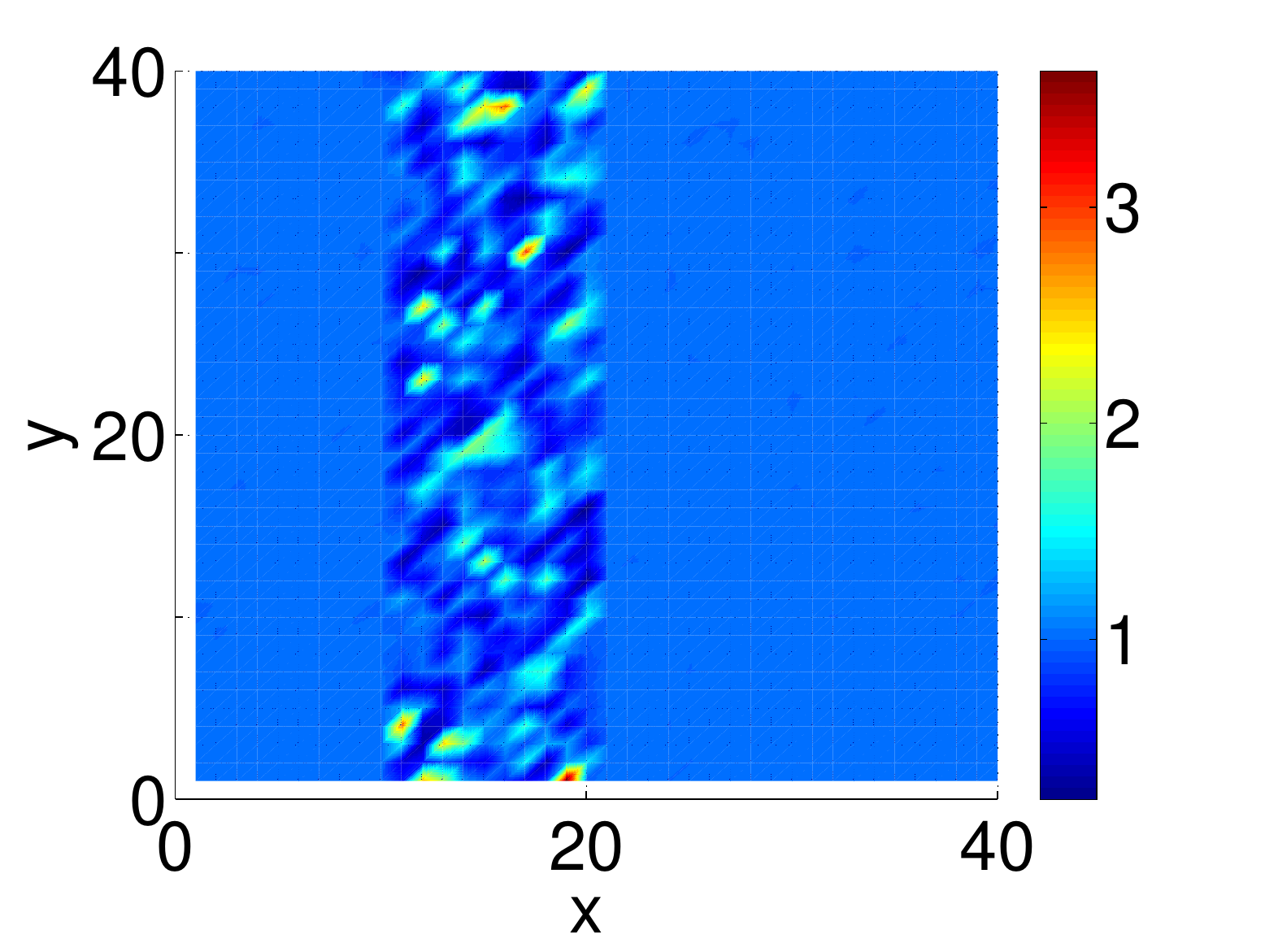} \label{fig:optobs_VAR_optvar} }
 \subfigure[4D-Var Analysis]{ \includegraphics[width=0.31\textwidth,height=0.25\textwidth]{reftraj0_lg} \label{fig:optobs_VAR_optrean}  }
 \caption{The minimization of the verification cost function, the optimized $h$ observation error covariances, and the resulting 4D-Var analysis
 using the improved values.}
 \label{fig:optobs_OBSVAR}
\end{figure}

Five iterations of the numerical optimization solver L-BFGS are performed, and the decrease in the values of $\Psi$ 
are plotted in Figure \ref{fig:optobs_VAR_cost}.
The solver converges monotonically and achieves a reduction of the verification cost function by a factor of six.
The 4D-Var analysis (initial time) is also improved by using the new data weights, as can be shown in Table \ref{Table:obsopt_gain2}.
The optimized values for the observation error variances are plotted in Figure \ref{fig:optobs_VAR_optvar}
(each variance is at the location of the corresponding sensor).
The tuned values are considerably larger in the area where observations were perturbed more significantly.
and thus the information with high noise levels is given a considerably reduced weight.
This result is significant because it shows the possibility to tune the data weights such as to
pick up the structure of observational noise, without having direct knowledge of the noise levels.

The analysis with initial error covariances is shown in Figure \ref{fig:optobs_VAR_init},
and the analysis with optimized weights is plotted in Figure \ref{fig:optobs_VAR_optrean}.


\begin{table}
\caption{Error norm of the 4D-Var reanalyzed initial solution before and after optimizing the observation weights.}
\centering
{
\begin{tabular}{|c||c|c|}
  \hline
  &  Initial weights & Optimized  weights \\
 \hline\hline
$\Vert \xa_0 - \x_0^{\rm reference} \Vert$ &  $0.4915$ & $0.0985$\\
 \hline
\end{tabular}
}
\label{Table:obsopt_gain2}
\end{table}

\subsection{Optimization of observation locations}

We now apply the methodology developed in Section \ref{sec:optimize-locations} to optimize the spatial configuration
of the sensor network such as to improve the performance of the data assimilation system. The approximate
gradients \eqref{eqn:dJdL-approximate} are used in the numerical calculations.

This experiment is set up slightly different. 
The observation network is sparse and consists of $300$ sensors measuring the fluid height $h$.
The physical locations of the sensors are specified by a given initial layout; each sensor location is described by
its Cartesian coordinates, which can vary continuously within the domain.
We change the coordinates of each sensor, i.e., the spatial configuration of the sensor network,
by solving the optimization problem \eqref{eqn:optimal-loc}. The minimization of the verification error
translates into an improved performance of the data assimilation system after relocating the observations.

\begin{figure}
\setcounter{subfigure}{0}
\centering
 \subfigure[Initial locations]{ \includegraphics[width=0.3\textwidth, trim=50 0 70 15, clip]{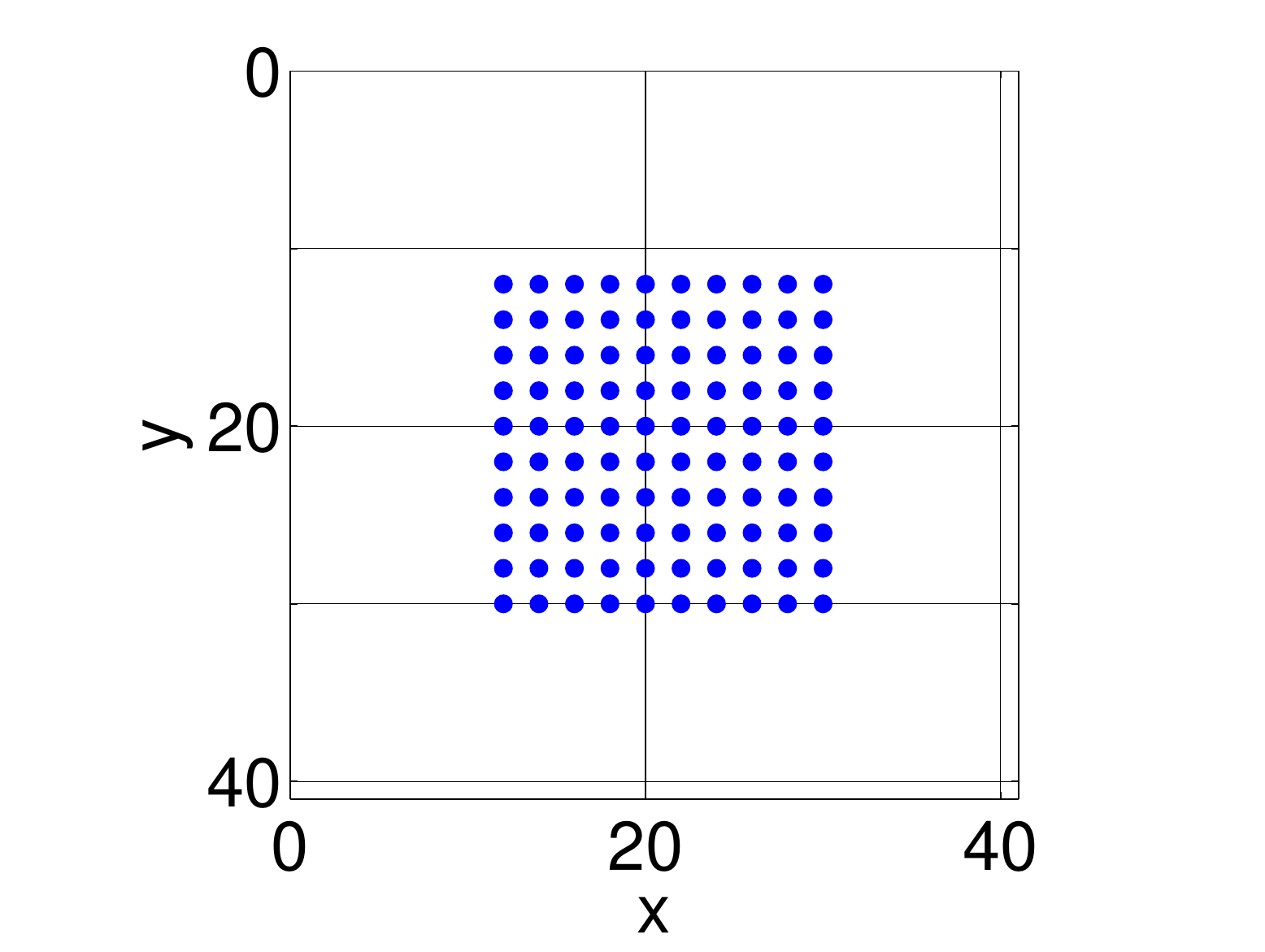} \label{fig:LOC_2x_initial} }
 \subfigure[L-BFGS convergence]{ \includegraphics[width=0.32\textwidth, height=0.3\textwidth, trim=5 0 5 10, clip]{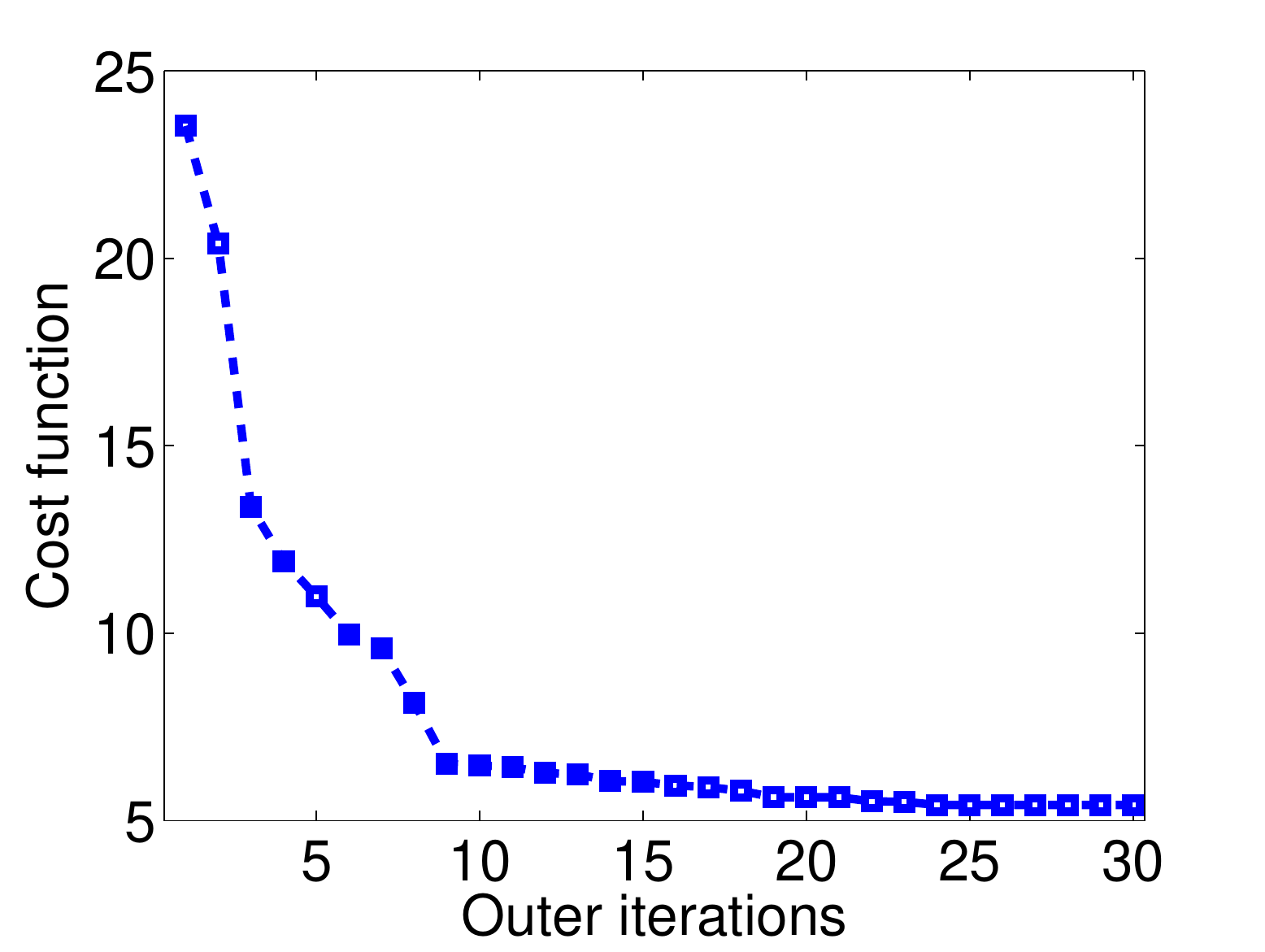} \label{fig:LOC_2x_cost} }
 \subfigure[Optimized locations]{ \includegraphics[width=0.3\textwidth, trim=50 0 70 15, clip]{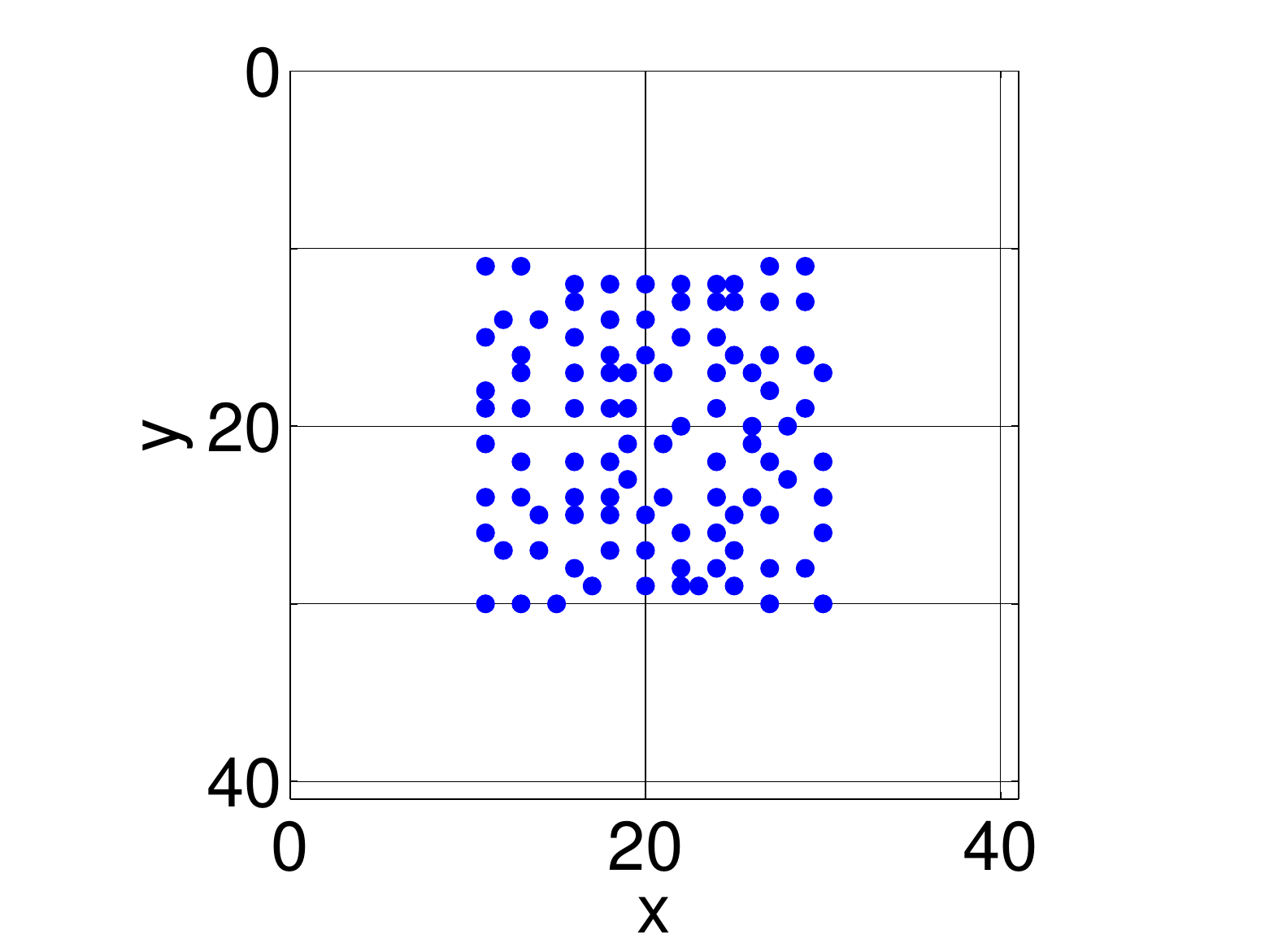} \label{fig:LOC_2x_optimal} }
\caption{The optimization of sensor locations for the first testing scenario: initial locations, numerical solver convergence, and optimal locations.}
\end{figure}

\begin{figure}
\setcounter{subfigure}{0}
\centering
 \subfigure[Initial locations]{ \includegraphics[width=0.3\textwidth, trim=50 0 70 15, clip]{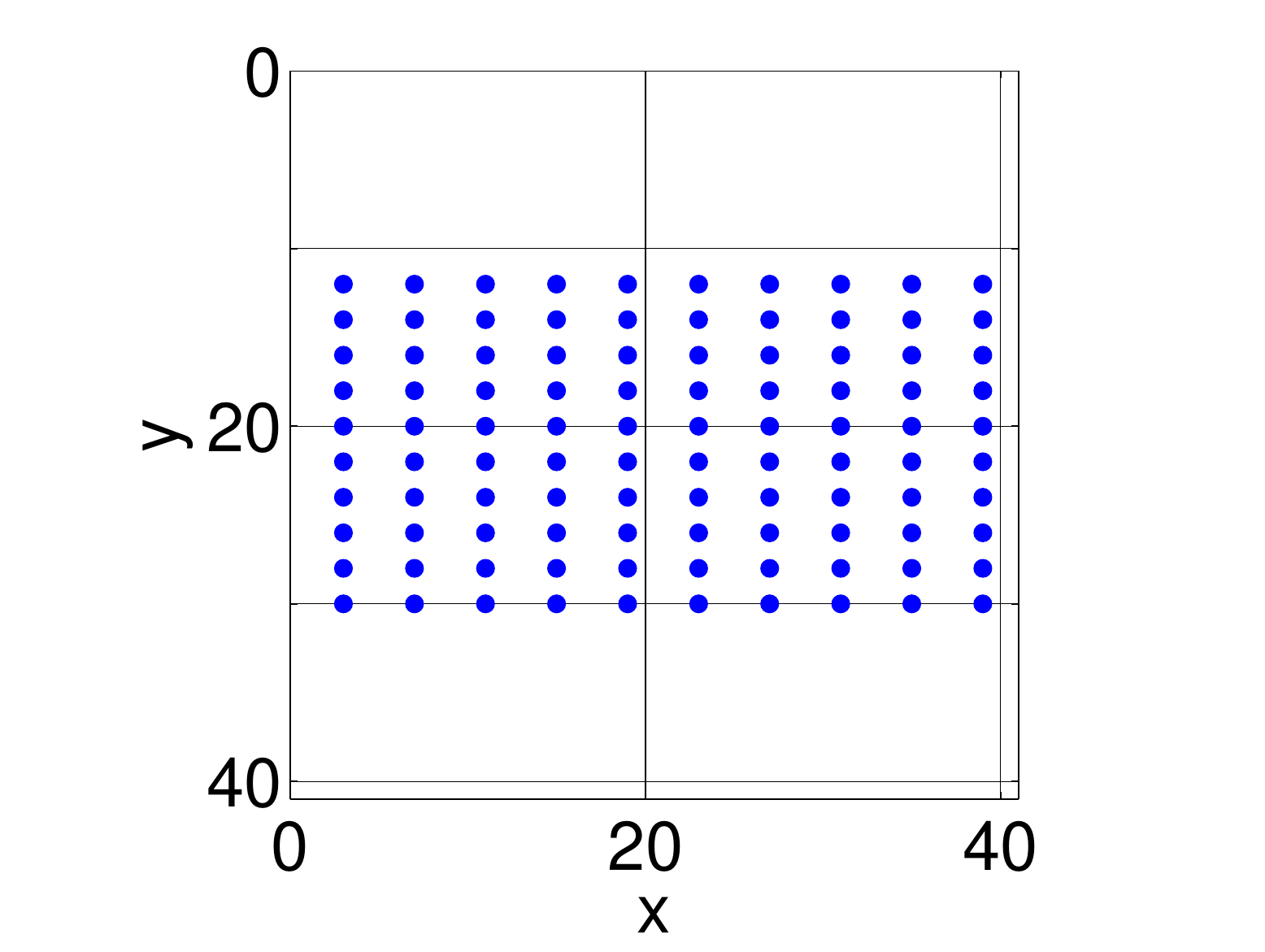} \label{fig:LOC_3x_initial} }
 \subfigure[L-BFGS convergence]{ \includegraphics[width=0.32\textwidth, height=0.3\textwidth, trim=5 0 5 10, clip]{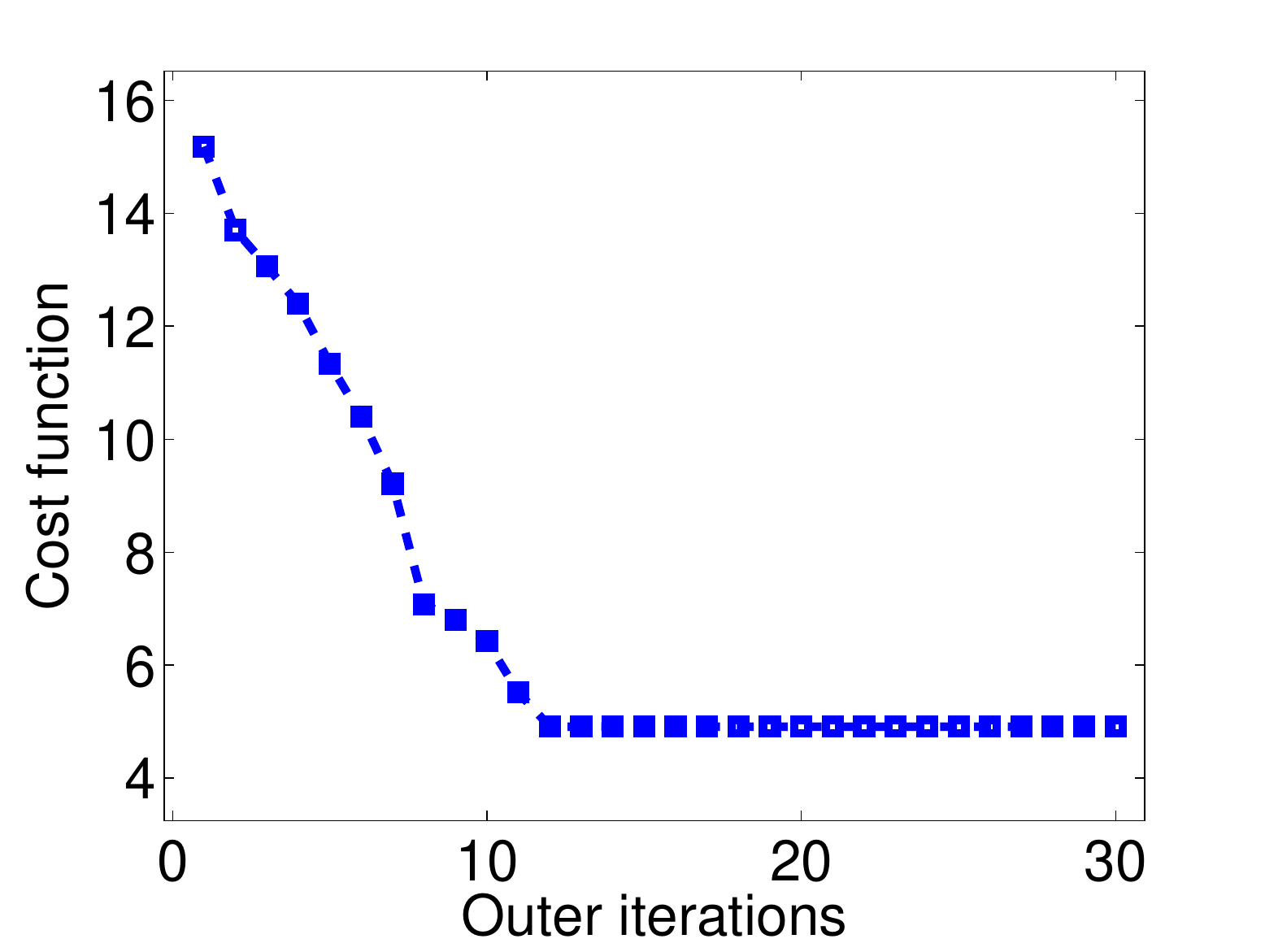} \label{fig:LOC_3x_cost} }
 \subfigure[Optimized locations]{ \includegraphics[width=0.3\textwidth, trim=50 0 70 15, clip]{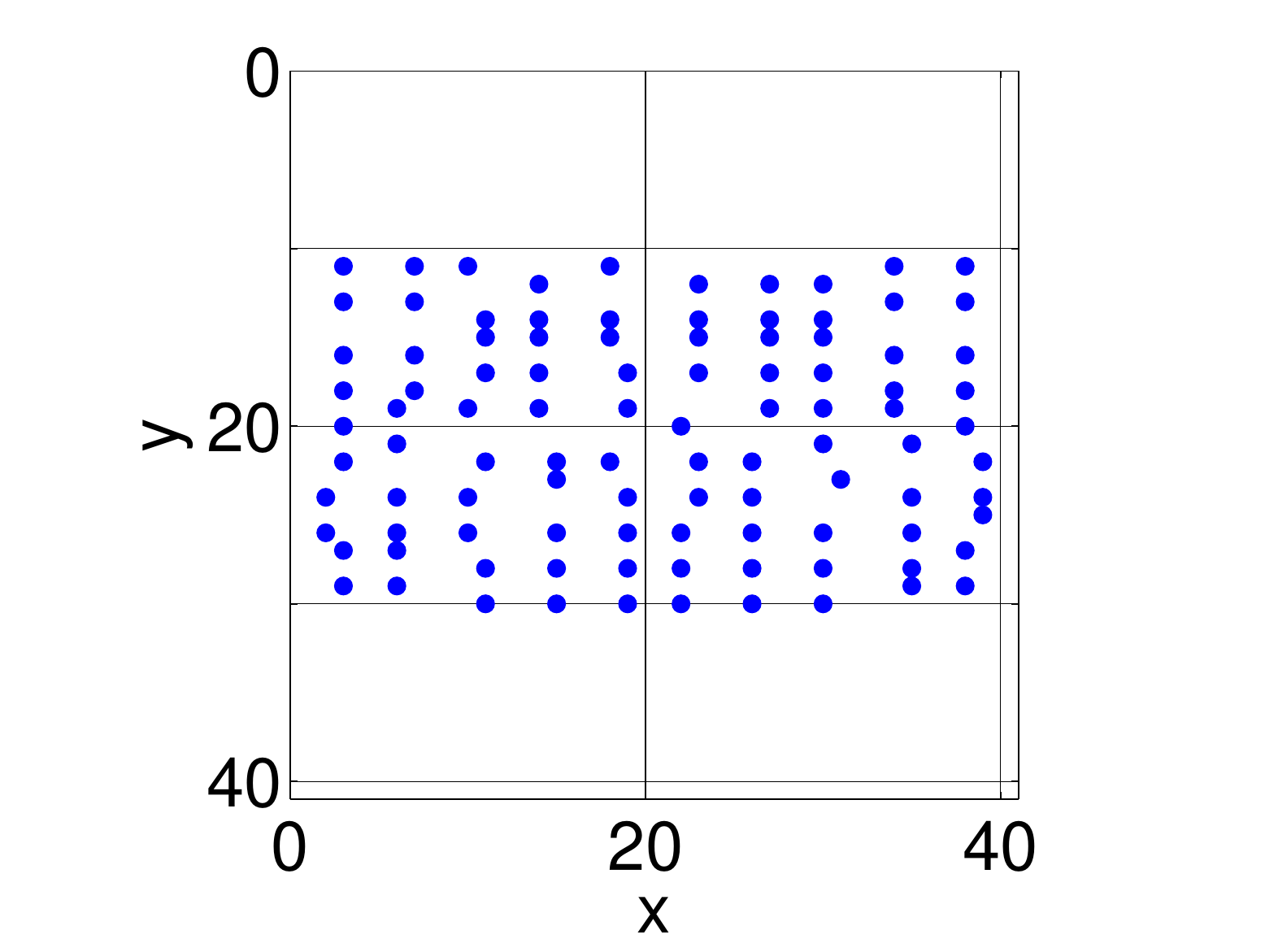} \label{fig:LOC_3x_optimal} }
\caption{The optimization of sensor locations for the second testing scenario: initial locations, numerical solver convergence, and optimal locations.}
\end{figure}

\begin{figure}
\setcounter{subfigure}{0}
\centering
 \subfigure[Initial locations]{ \includegraphics[width=0.3\textwidth, trim=50 0 70 15, clip]{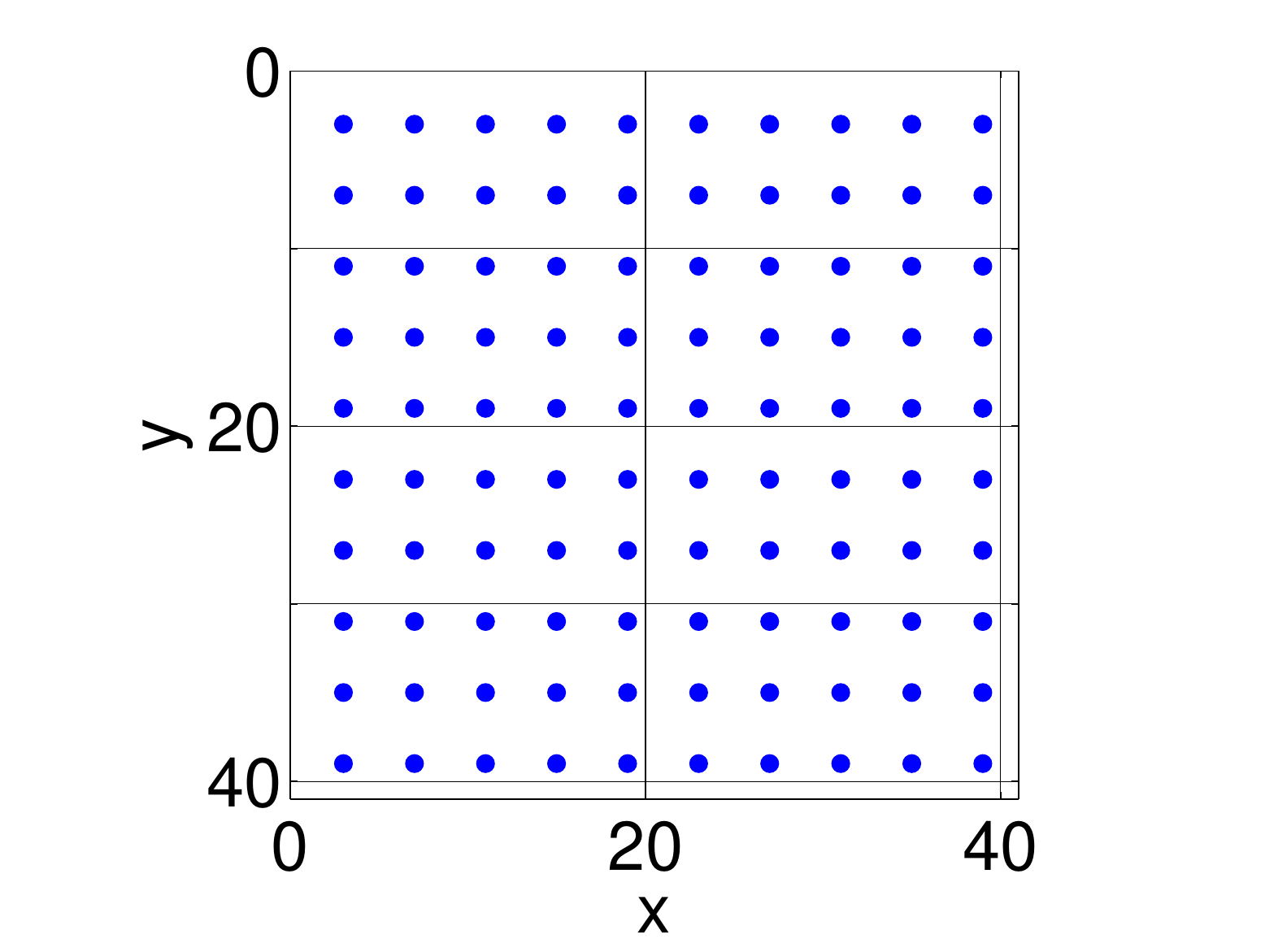} \label{fig:LOC_4x_initial} }
 \subfigure[L-BFGS convergence]{ \includegraphics[width=0.32\textwidth, height=0.3\textwidth, trim=5 0 5 10, clip]{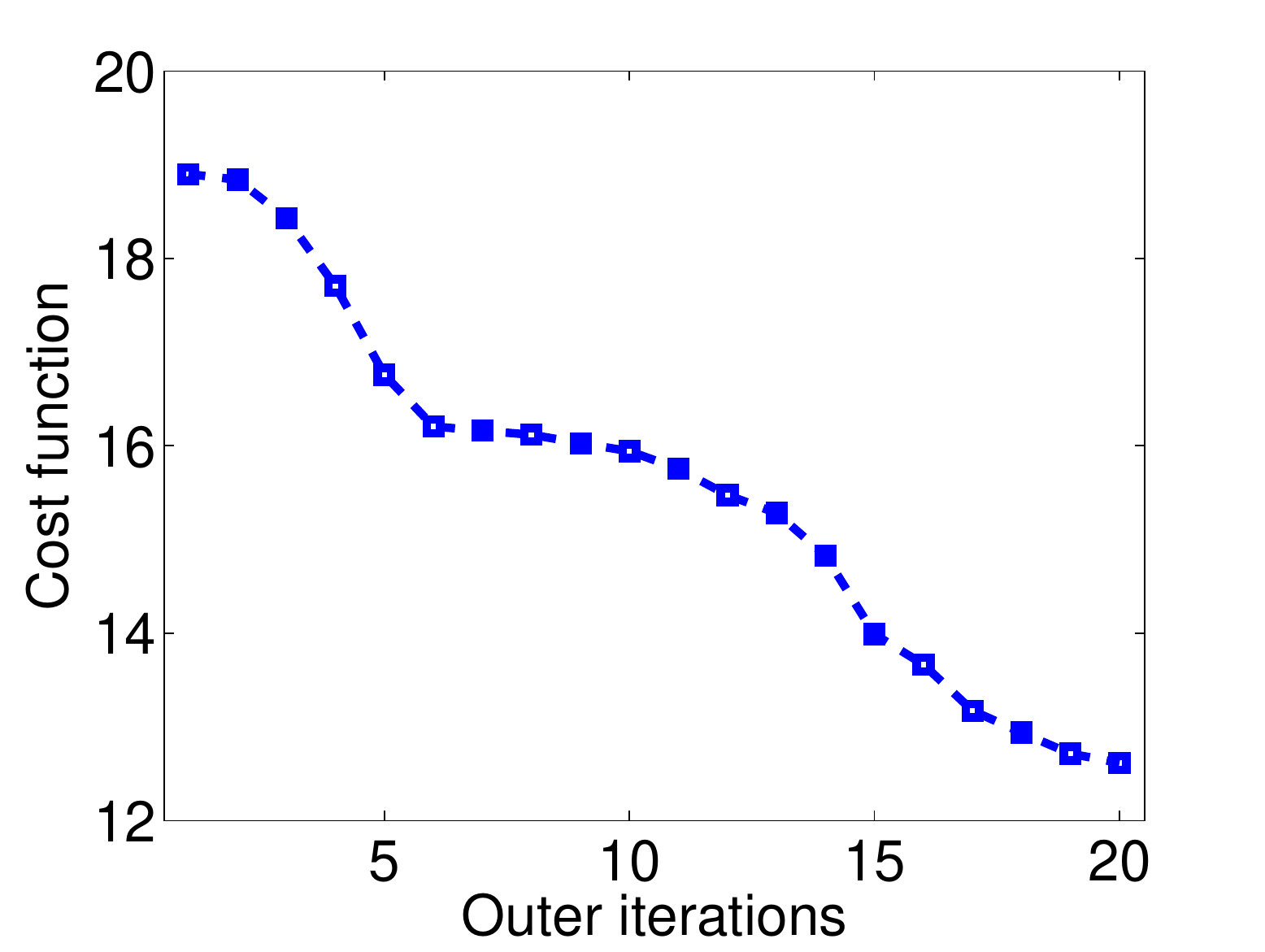} \label{fig:LOC_4x_cost} }
 \subfigure[Optimized locations]{ \includegraphics[width=0.3\textwidth, trim=50 0 70 15, clip]{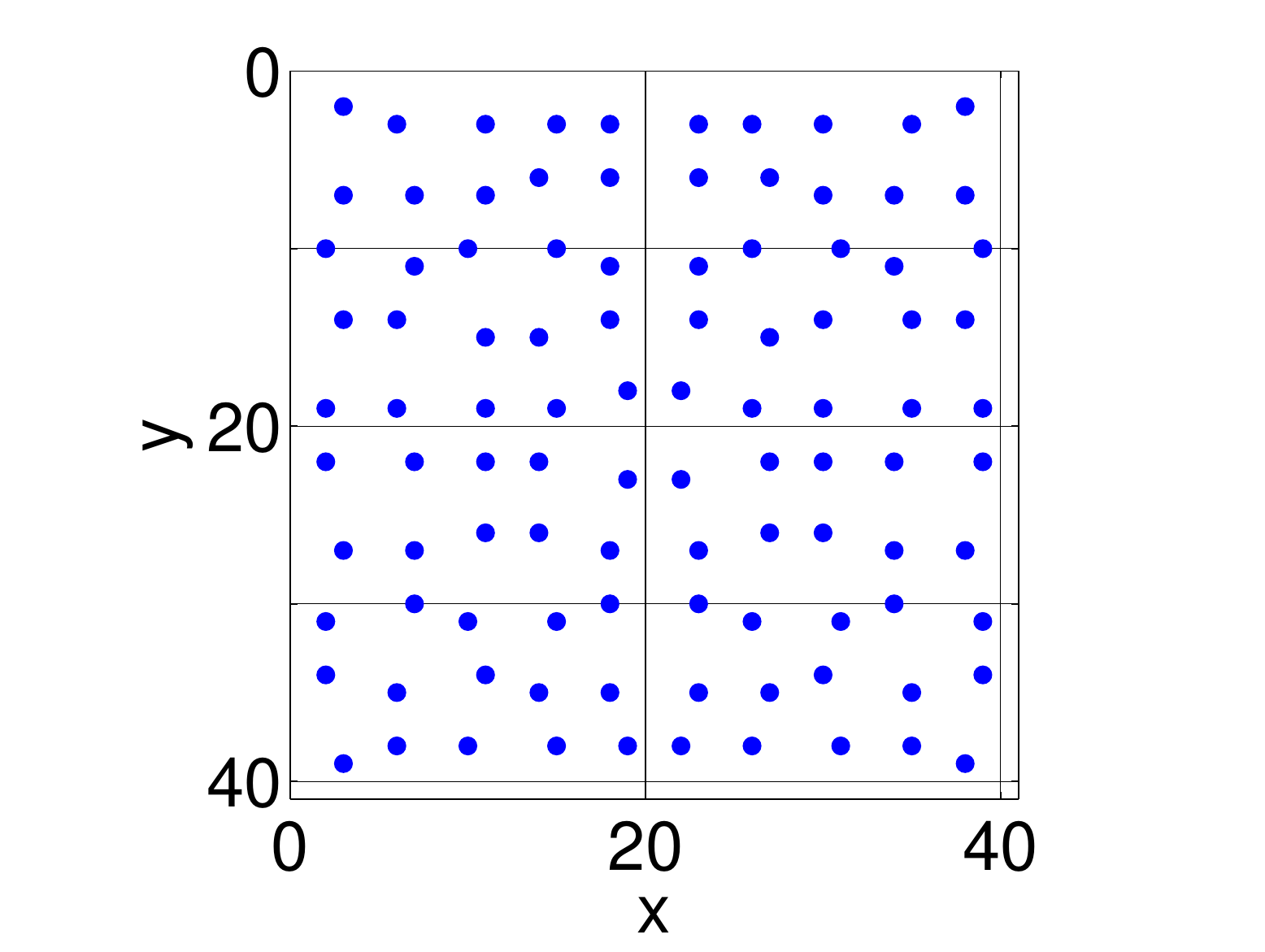} \label{fig:LOC_4x_optimal} }
 \caption{The optimization of sensor locations for the third testing scenario: initial locations, numerical solver convergence, and optimal locations.}
 \label{fig:optobs_OBSLOC}
\end{figure}

The optimization problem is solved for three scenarios, distinguished by the initial sensor locations.
Each scenario starts from an equidistant spatial distribution of sensors, as shown in 
Figures \ref{fig:LOC_2x_initial}, \ref{fig:LOC_3x_initial}, and \ref{fig:LOC_4x_initial}, respectively.
L-BFGS is ran for solving \eqref{eqn:optimal-loc} over $30$ iterations, and the corresponding 
decreases in cost function values for each scenario are plotted in 
Figures \ref{fig:LOC_2x_cost}, \ref{fig:LOC_3x_cost}, and \ref{fig:LOC_4x_cost}, respectively.
We can notice that for the first two scenarios the cost function decrease is converging to a minimum.
Meanwhile, the iterative solver broke down for the third scenario after decreasing the cost function within the first $20$ iterations.

The resulting optimal locations for each scenario are shown in 
Figures \ref{fig:LOC_2x_optimal}, \ref{fig:LOC_3x_optimal}, and \ref{fig:LOC_4x_optimal}, respectively. 

In each scenario the optimization adjusts slightly the locations of the sensors; this is sufficient
for obtaining a considerable reduction in the verification cost function $\Psi$, i.e.,
for obtaining a measurable improvement in the data assimilation system performance.
Table \ref{Table:obsopt_gain3} contains the true error norm of the 4D-Var reanalyzed initial solution
for each solution and we can notice corresponding improvements.


%

While the performance is improved in each of the three scenarios, the computed solutions 
are clearly only local minima of \eqref{eqn:optimal-loc}, located in the vicinity of the corresponding
initial configurations. Note that the global optimum coincides for all three cases, and represents the best
locations of $300$ sensors within the domain.
The convergence to a local optimum is to be expected for the quasi-Newton
approach used herein, as the gradients mostly contain information that is locally valid.



\begin{table}
\caption{Error norm of the 4D-Var reanalyzed initial solution before and after optimizing sensor locations.}
\centering
{
\footnotesize
\begin{tabular}{|c|c||c|c|}
  \hline
  Assimilation Scenario & &  Initial locations & Optimized locations \\
 \hline\hline
1 &  & $0.3923$ & $0.1960$\\
 \cline{1-1} \cline{3-4}
2 & $\Vert \xa_0 - \x_0^{\rm reference} \Vert$ & $0.3213$ & $0.1906$\\
 \cline{1-1} \cline{3-4}
3 &  &  $0.3327$ & $0.2897$\\
 \hline
\end{tabular}
}
\label{Table:obsopt_gain3}
\end{table}


\section{Conclusions and Future Work}\label{sec:optobs_concl}

This paper, develops a computational framework for improving the performance of
4D-Var data assimilation systems. The approach is based on
optimizing various  4D-Var parameters with respect to a forecast aspect quantified by 
a verification cost functional. The particular aspect considered here is the norm of a forecast error
estimate. System parameters include the observation values, the observation covariances, and the location of
sensors.

The verification functional is defined on the analysis, i.e., on the solution of the data assimilation problem.
We formulate a constrained continuous optimization problem to find the best parameter values
for the data assimilation system. The cost function is the verification functional,
and the constraints are the optimality conditions of the 4D-Var system. Thus we have an
``optimization-constrained optimization problem''.
The computational solution employs gradient-based numerical optimization methods.
The gradient of the forecast aspect of interest with respect to data assimilation system parameters
is computed (relatively) efficiently using adjoint models and the framework of 4D-Var sensitivity analysis. 
The proposed methodology is the first of its kind and can be readily integrated in 
data assimilation studies for weather, climate, air-quality and many others.

Our study shows that an optimization problem constrained by the 4D-Var problem itself is tractable
and can improve the quality of analyses generated by the data assimilation system.
Numerical results obtained with a test shallow water equations system illustrate how
the proposed approach successfully optimized the values 
of observations, observation error covariances, and sensor locations.

This research opens the path for solving a multitude of problems related to the
optimal configuration of specific 4D-Var data assimilation systems. 
Future work will extend the optimization procedure to other parameters of the data assimilation system
such as the background covariance and, in the context of weakly constrained 4D-Var, the model error
covariances. The approach will benefit from developing more efficient techniques
to compute the derivatives and improving the convergence of the optimization process.
We plan to consider alternative formulations of the verification cost functional, such as ones based
on information theory, and to extend the framework to include mixed integer optimization problems.
This will extend the capabilities of the framework to optimize the observing network by turning sensors on or off,
or by selecting the best alternatives from a given set of feasible locations.

\section*{Acknowledgements}

This work was supported by the National Science Foundation through the awards NSF DMS-0915047, NSF CCF-0635194, NSF CCF-0916493 and NSF OCI-0904397,
and by AFOSR DDDAS program through the awards FA9550--12--1--0293--DEF and AFOSR 12-2640-06.


\newpage
\setcounter{page}{1}

\bibliography{optobs}
\bibliographystyle{unsrt}

\end{document}

%% file: sandu_template_symbols.tex
\newcommand{\Jfunc}{\mathcal J }
\newcommand{\Model}{\mathcal M}
\newcommand{\M}{\mathbf{M}}

\newcommand{\Hobs}{\mathcal{H}}
\newcommand{\HH}{\mathbf H}

\newcommand{\B}{\mathbf{B}}
\newcommand{\C}{\mathbf{C}}
\newcommand{\R}{\mathbf{R}}

\newcommand{\x}{   \mathbf{x} }
\newcommand{\xb}{ \mathbf{x}^{\rm b} }
\newcommand{\xa}{ \mathbf{x}^{\rm a} }

\newcommand{\xv}{ \mathbf{x}^{\rm v} }
\newcommand{\y}{ \mathbf{y} }

\newcommand{\z}{ \mathbf{z} }

\renewcommand{\u}{ \mathbf{u} }

\renewcommand{\H}{\mathcal{H}}



%% file: logo.tex
\thispagestyle{empty}
\setcounter{page}{0}

\begin{Huge}
\begin{center}
Computational Science Laboratory Technical Report CSL-TR-4-2013 \\
\today
\end{center}
\end{Huge}
\vfil
\begin{huge}
\begin{center}
Alexandru Cioaca and Adrian Sandu
\end{center}
\end{huge}

\vfil
\begin{huge}
\begin{it}
\begin{center}
An Optimization Framework to Improve \\ 
4D-Var Data Assimilation System Performance
\end{center}
\end{it}
\end{huge}
\vfil

\begin{large}
\begin{center}
Computational Science Laboratory \\
Computer Science Department \\
Virginia Polytechnic Institute and State University \\
Blacksburg, VA 24060 \\
Phone: (540)-231-2193 \\
Fax: (540)-231-6075 \\ 
Email: \url{sandu@cs.vt.edu} \\
Web: \url{http://csl.cs.vt.edu}
\end{center}
\end{large}

\vspace*{1cm}

\begin{tabular}{ccc}
\includegraphics[width=2.5in]{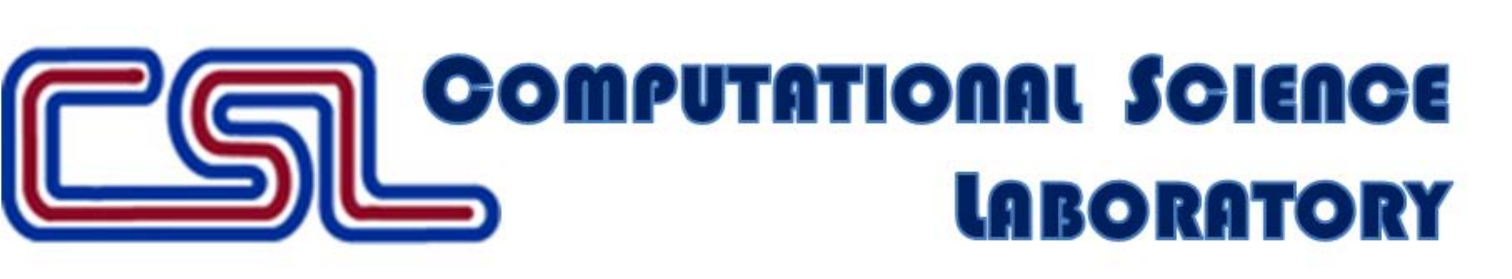}
&\hspace{2.5in}&
\includegraphics[width=2.5in]{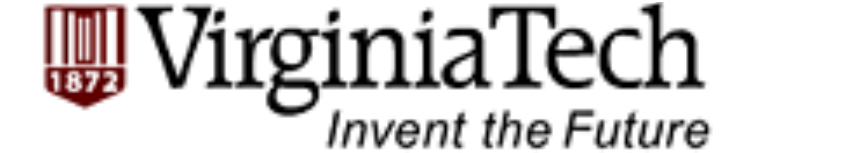} \\
{\bf\em Innovative Computational Solutions} &&\\
\end{tabular}

\newpage